\documentclass[a4paper,twocolumn]{article}
\usepackage[margin=0.6in]{geometry}
\usepackage[T1]{fontenc}
\usepackage[utf8]{inputenc}
\usepackage{lmodern}
\usepackage[english]{babel}

\usepackage{graphicx}
\usepackage{amssymb}
\usepackage{overpic}

\usepackage{hyperref}
\hypersetup{
    colorlinks=true,
    linkcolor=black,
    citecolor=black,
    filecolor=black,
    urlcolor=black,
    linkbordercolor = {1 1 1},
}

\usepackage{algpseudocode}
\usepackage{algorithm}

\algblockdefx{Step}{EndStep}[1]{#1}{}
\algtext*{EndStep}

\usepackage[square,numbers]{natbib}

\title{Addressing the ``minimum parking'' problem for on-demand mobility}

\author{
	D\'aniel Kondor$^{1,\ast}$, Paolo Santi$^{2,3}$, Diem-Trinh Le$^{1}$,\\[0.5ex]
	Xiaohu Zhang$^{1,2}$, Adam Millard-Ball$^{4}$, Carlo Ratti$^{1,2}$ \\[1ex]
		\normalsize{$^1$Singapore-MIT Alliance for Research and Technology, Singapore}\\[0.3ex]
		\normalsize{$^2$Senseable City Laboratory, MIT, Cambridge MA 02139 USA}\\[0.3ex]
		\normalsize{$^3$Istituto di Informatica e Telematica del CNR, Pisa, Italy}\\[0.3ex]
		\normalsize{$^4$University of California Santa Cruz, Santa Cruz CA, USA}\\[0.3ex]
		\normalsize{$^\ast$ E-mail: \texttt{dkondor@mit.edu}}}
\date{\today}
\begin{document}

\maketitle

{\bf
Parking infrastructure is pervasive and occupies large swaths of land in cities~\cite{Manville2005}. However, on-demand (OD) mobility -- such as commercial services Uber, Grab or Didi -- has started reducing parking needs in urban areas around the world~\cite{Henao2019}. This trend is expected to grow significantly with the advent of autonomous driving~\cite{Fagnant2015,Zhang2017,paper1}, which might render on-demand mobility predominant~\cite{Burns2013,Martinez2017,Stead2019}. Recent studies have started looking at expected parking reductions with on-demand mobility~\cite{Zhang2017,Xu2017}, but a  systematic framework is still lacking. In this paper, we apply a data-driven methodology based on shareability networks~\cite{Santi2018,Santi2014} to address what we call the ``minimum parking'' problem: what is the minimum parking infrastructure needed in a city for given on-demand mobility needs?
While solving the problem, we also identify a critical tradeoff between two public policy goals: less parking means increased vehicle travel from deadheading between trips. By applying our methodology to the city of Singapore we discover that parking infrastructure reduction of up to 86\% is possible, but at the expense of a 24\% increase in traffic measured as vehicle kilometers travelled (VKT).
However, a more modest 57\% reduction in parking is achievable with only a 1.3\% increase in VKT.
We find that the tradeoff between parking and traffic obeys an inverse exponential law which is invariant with the size of the vehicle fleet, leading to a simple methodology to estimate aggregate parking demand in a city. Finally, we analyze parking requirements due to passenger pick-ups and show that increasing convenience produces a substantial increase in parking for passenger pickup/dropoff. The above mathematical findings can inform policy-makers, mobility operators, and society at large on the tradeoffs required in the transition towards pervasive on-demand mobility.
}

	Cities currently devote a large amount of space and resources to provide parking, primarily used by private cars that are idle 95\% of the time~\cite{Burns2013}. For example, in Los Angeles County, where there are 3.3 parking spaces per car, the total area of parking spaces is equal to 14\% of total incorporated land area and is 1.4~times larger than the total area used by roads~\cite{Chester2015}. In dense city centers, parking can account for an even larger share -- the total floor area dedicated to parking is between 25\% and 81\% of land area~\cite{Manville2005} and can be larger than the floor area of office or retail use that it serves~\cite{Shoup1995}.
	Using floor area as a metric counts each level of a parking garage separately, and so the spatial footprint of parking is less, but it can still account for over 5\% of total urban land areas~\cite{Davis2010a,Davis2010b}. Similar levels of parking are mandated in many Asian cities~\cite{Barter2011}.
	
	However, changes are starting to be apparent with the increased popularity of shared mobility services provided by on-demand vehicle (OV) fleets, such as ridehailing services, which can increase vehicle on-road time and reduce parking needs~\cite{Henao2019}. Further opportunities are foreseen with the gradual transition to autonomous vehicles (AVs)~\cite{Fagnant2015,Gruel2016}, which are expected to reduce the number of privately-owned cars and further popularize shared mobility~\cite{Bansal2017,Daziano2017} due to being more cost-effective than both taxis and private vehicles~\cite{Burns2013,Brownell2014,Bosch2018}. OV fleets could reduce a city's parking needs through several mechanisms. First, thanks to vehicle and/or ride sharing, OVs are expected to reduce the size of the vehicle fleet by 40\% to 90\%~\cite{Fagnant2014,Fagnant2015b,Burns2013,Spieser2014,Santi2018}, accompanied by similar reduction in the demand for parking~\cite{Zhang2015,Zhang2017,paper1}. Furthermore, OVs have no need to park at their destination, and can return home, park remotely, or even cruise (circle) around~\cite{Harper2018,MillardBall2019}, resulting in increased utilization of a smaller amount of parking.
	AVs can further reduce the spatial footprint of parking facilities by exploiting better maneuvering capabilities and the fact that individual vehicles need not be accessible to humans when parked~\cite{Timpner2015,Kong2018,Nourinejad2018}. While autonomous mobility is till forthcoming, most of the benefits related to the use of shared, on-demand mobility could be realized today. For this reason, in this study we focus on fleets of OVs that can be either autonomous or chauffeured, and measure parking infrastructure as number of parking spots needed at the city scale, without accounting for possible additional benefits provided by autonomous vehicles mentioned above.
	
	\begin{figure*}
		\centering
		\includegraphics[width=3.2in]{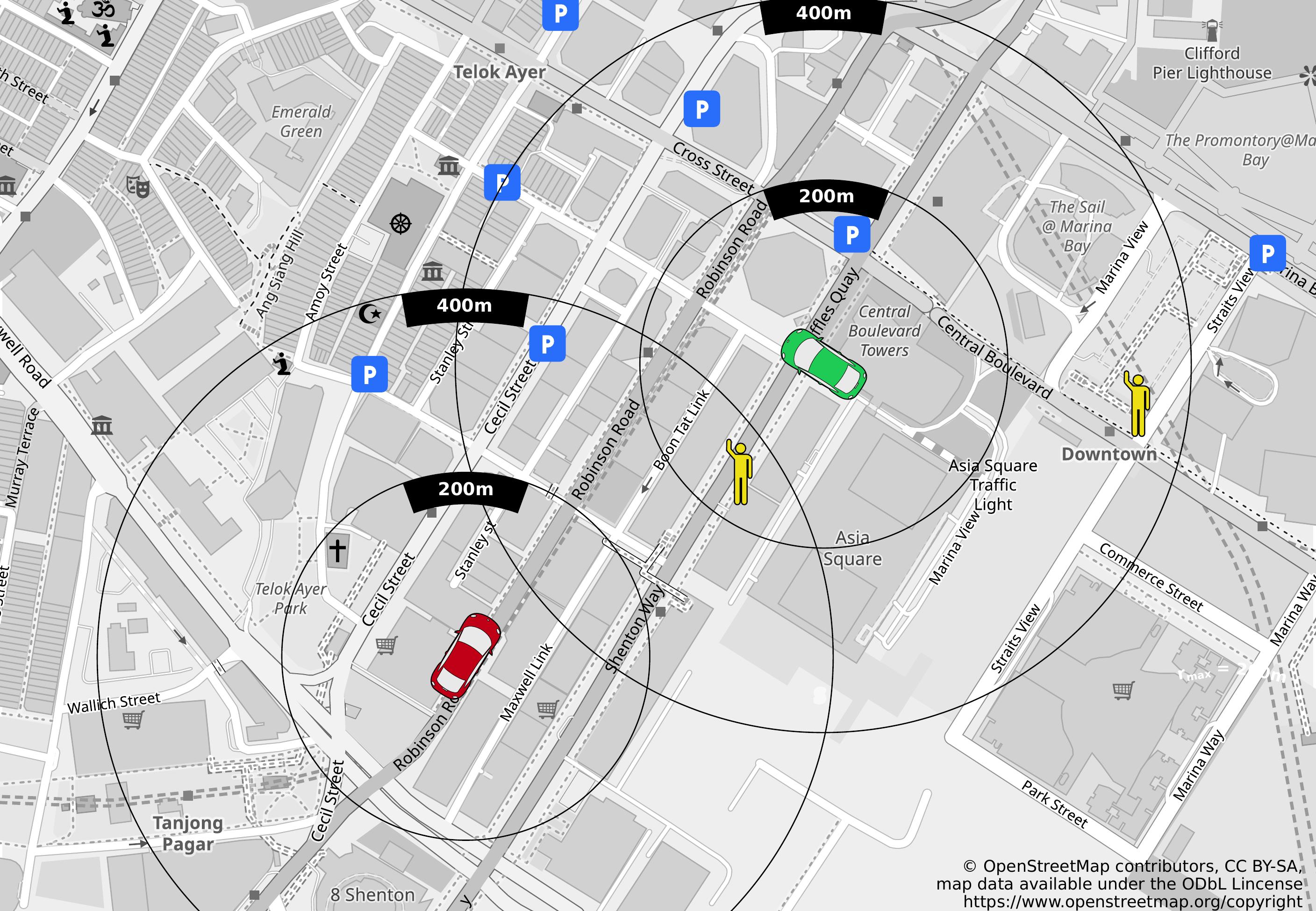} \quad
		\includegraphics{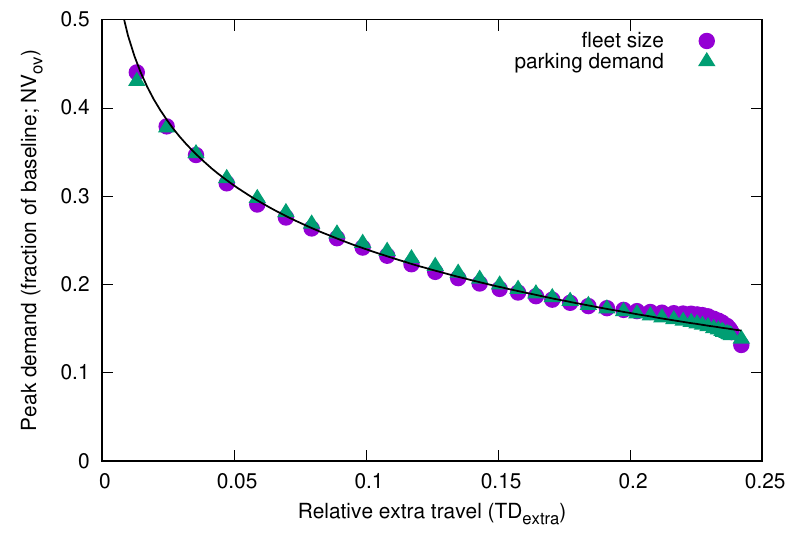}
			\caption{Tradeoff between parking demand and vehicle travel. The left panel shows an illustration of the problem and the process used to estimate parking requirements. It shows a simplified case with two on-demand vehicles, several available parking spaces and two passengers requesting a ride: the red vehicle is just dropping off a passenger and would need to find parking or a next passenger, while the green vehicle is already parked and would be available for new trips. We show search radiuses of $r_\mathrm{max} = 200\,\mathrm{m}$ and $400\,\mathrm{m}$ for each vehicle. For the smaller radius, the red vehicle cannot find any parking, while the green vehicle can only serve one of the passengers. In this case, we would need to add further parking for the red vehicle and one additional vehicle to serve the passenger farther away. For the larger radius, the red vehicle has a choice of parking or serving one passenger, while the green vehicle could serve either passenger. Performing a maximum matching in this case will assign the two vehicles to the two passengers, thus resulting in a solution with less vehicles and parking, but more extra travel. Note that $r_\mathrm{max}$ in this figure is shown as Euclidean distance for illustration purposes only; in our actual analysis, we use distances calculated along the road network. On the right panel, we show estimation results with different values of $r_\mathrm{max}$, ranging from $500\,\mathrm{m}$ to unlimited. We display relative demand for parking and vehicle fleet size as a function of increased VKT. The fitted line suggests that increase in VKT grows exponentially as a function of decrease in fleet size and parking requirements (see SI for discussion). }
		\label{newres1}
	 \end{figure*}

	Despite the initial studies mentioned above, a precise mathematical solution to the minimum parking problem is still missing. We can phrase the problem as follows: given a number $NT$ of private vehicle trips, what is the minimum parking infrastructure needed to support this demand? In this paper, we use shareability networks and related optimal dispatching algorithms~\cite{Santi2018,paper1} and derive the minimum number of parking spots $NP_{ov}$ needed in a city, for a given size $NV_{ov}$ of the fleet used to serve the $NT$ trips. Clearly, on demand vehicles could be moving all the time, reducing the number of parking spots to zero. Hence, we introduce a parameter $r_\mathrm{max}$ that bounds the number of empty kilometers a vehicle can drive in-between trips, and show how $NP_{ov}$ varies as a function of $r_\mathrm{max}$. Finally, we measure the total additional travel distance $TD_{extra}$ needed to serve the $NT$ trips with $NV_{ov}$ vehicles and $NP_{ov}$ parking spots, and formally characterize a tradeoff between the three quantities at stake (number of vehicles, parking infrastructure, and traveled distance) in a case study of the city of Singapore.
		
	For the present study, we need to capture overall vehicular mobility demand in Singapore. We use data from SimMobility, probably the most precise and comprehensive simulator for urban mobility which incorporates a detailed model of people’s movements in Singapore~\cite{Adnan2016}. We concentrate on the trips made in private vehicles and investigate the scenario in which all of these trips are served by on-demand vehicles instead. We use a methodology based on bipartite matching of vehicles to trips and parking spaces to arrive at an estimate of $NV_{ov}$, $NP_{ov}$, and $TD_{extra}$~\cite{paper1,Santi2018} -- see Fig.~\ref{newres1} (left panel) for an illustration and the Materials and Methods and Supplementary Material for a detailed description. We compare our results to an estimation of parking supply based on land-use data, parking requirements and constraints based on the SimMobility trip dataset.
		
	The right panel in Figure~\ref{newres1} reports the main results of the analysis. We find that parking could be reduced by as much as 86\% from the current estimate of around 1.37 million to about 189 thousand. At the same time, the number of vehicles would be reduced by a similar ratio, from around 676 thousand to below 89 thousand.
	
	\begin{figure*}
		\centering
		\includegraphics{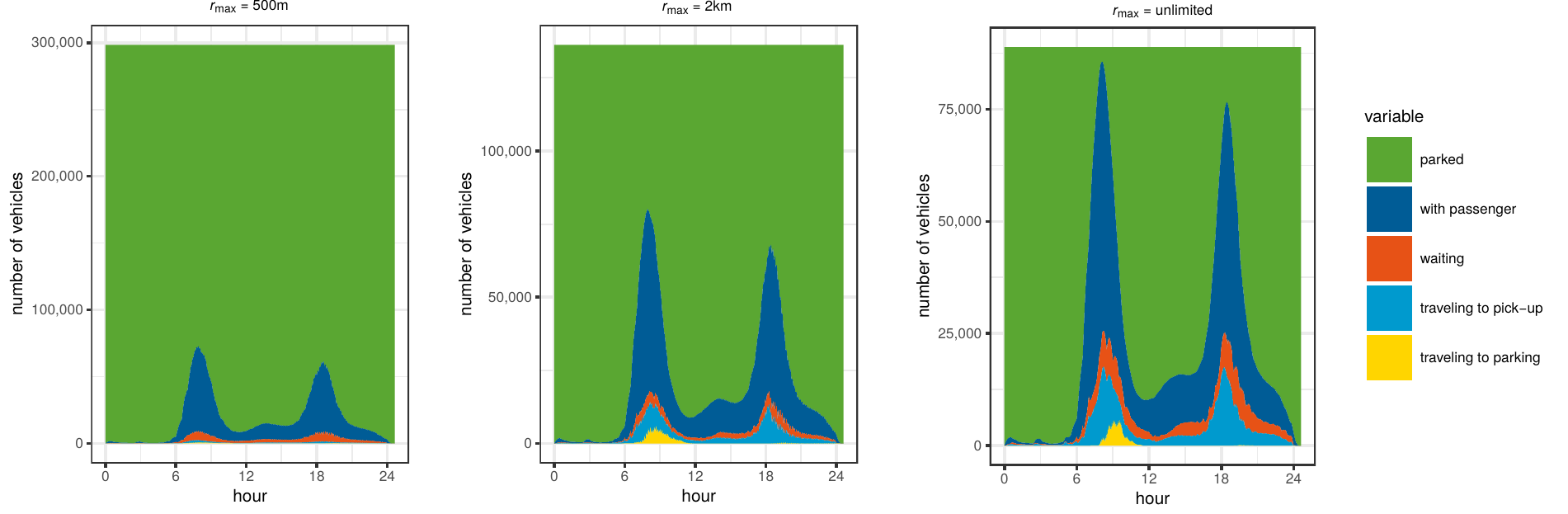}
		\caption{Utilization of fleet during the day for $r_\mathrm{max} = 500\,\mathrm{m}$ (left), $2\,\mathrm{km}$ (middle) and
			unlimited (right). The upper limit of the y-axis scale indicates the total fleet size necessary, ranging from $\sim$300,000 vehicles in the left panel to $\sim$100,000 in the right panel.}
		\label{fleet_utilization}
	\end{figure*}
		
	However, savings in parking come at a cost: in our minimum parking scenario, the OV fleet has about 24.2\% more travel (total VKT) than the baseline case of people making the same trips in private vehicles\footnote{Note that this estimate relies on the assumption that private car users can find parking exactly at their destination which is not true in all cases. Thus, we see our estimate of VKT increase as an upper bound.}. This highlights a tradeoff between parking reduction and increased vehicle travel, that we control via the parameter $r_\mathrm{max}$, that determines the maximum distance OVs are allowed to travel to reach parking or the next passenger. While in practice, $r_\mathrm{max}$ will be a parameter controlled by economics of fleet operations or regulations set by governments, for the purpose of the current study, we treat it as a design parameter that allows us to explore the tradeoff between parking, fleet size and vehicle travel. If human mobility flows were perfectly balanced and evenly distributed throughout the day, no such tradeoff would occur -- i.e., if for each passenger delivered at a destination, there were another passenger at the same location available for immediate pickup. In practice, however, there is overwhelming literature showing that human flows are highly unbalanced spatially and temporally~\cite{Alexander2015,Azevedo2016,Diao2016,Halvorsen2019}; thus, it is important to observe that, unless sharing of \emph{rides} is considered, the sharing of \emph{vehicles} cannot directly reduce the total traveled distance (see the Supplementary Material, Section~2, Figs.~\ref{shareability} and~\ref{ridesharing} on how sharing of rides can offset VKT increase).
	
	In Figure~\ref{newres1} (right panel), each point represents one realization of our estimation with values of $r_\mathrm{max}$ ranging from $500\,\mathrm{m}$ to infinity. The first striking observation is that relative reductions in parking demand closely track relative reductions in fleet size. In fact, there is an almost constant ratio of $\sim 2$ between the two quantities (see SI Table~\ref{tab:ratios}). Intuitively, this is explained by the fact that each OV uses parking in two locations: close to residential areas where people start their commute in the morning, and in the city center where parking is necessary to be able to serve trip in succession (where moving to park in a remote location would take up too much time). Enforcing a strict separation between short-term and long-term parking can change this picture however, as we show in Supplementary Material (section~3.3, SI Fig.~\ref{ratios1}).
	
	A second important observation is that the relationship between the extra distance traveled $TD_{extra}$ and the unified variable $N_{ov}=(NV_{ov},NP_{ov})$ can be empirically described by an exponential function
	\[
	 TD_{extra} = e^{-a NV_{ov}}~,
	 \]
	where $a = 9.6$ ($R^2 = 0.976$). This result implies a law of diminishing return for the control parameter $r_\mathrm{max}$. A short search radius of $500~\mathrm{m}$ is already sufficient to absorb many inefficiencies related to asymmetric and unbalanced mobility flows, achieving over 57\% reductions in both fleet size and parking needs compared to the baseline scenario. At the same time, $TD_{extra}$ is limited to about 1.3\%. On the other hand, if higher reductions in fleet size/parking infrastructure are sought, the search radius should be significantly increased, up to $5~\mathrm{km}$ or above. Despite the 10-fold increase in search radius, $N_{ov}$ is reduced by only 82\% (2.45-fold reduction), while empty travel kilometers are increased to about 18\% over $TD_{cur}$ (over 13-fold increase).
	
	The significance of $r_\mathrm{max}$ is even more evident when looking at the utilization of the fleet during the day as displayed in Fig.~\ref{fleet_utilization}. For small values of $r_\mathrm{max}$, fleet usage is limited by the ability of vehicles to reach trip requests. For large values of $r_\mathrm{max}$, fleet size is determined by the need to serve peak hour demand which is significantly higher than at other times. 
	
	The analysis so far considers only \emph{parking} demand, i.e.~storage for OVs that are not in service. In the following, we further investigate \emph{pick-up} demand, i.e.~space for OVs to wait while picking up a passenger. If users prefer short or zero wait times for an on-demand trip, the vehicle will need to (at least in expectation) arrive prior to a trip's start time and wait in a suitable area. Ordering a vehicle in advance can ensure this, but can introduce an uncertainty in how long the vehicle needs to wait for the passenger.
	
\begin{figure*}[t]
		\centering
		\includegraphics{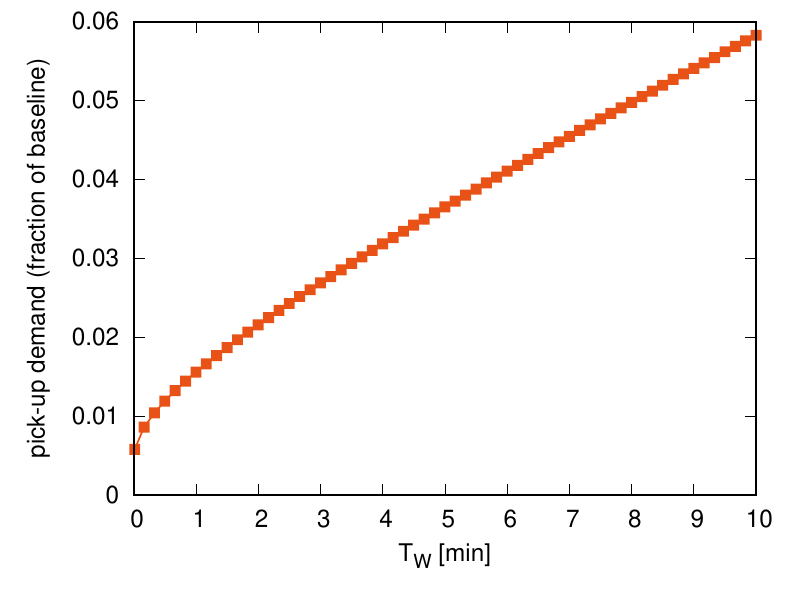}
		\includegraphics{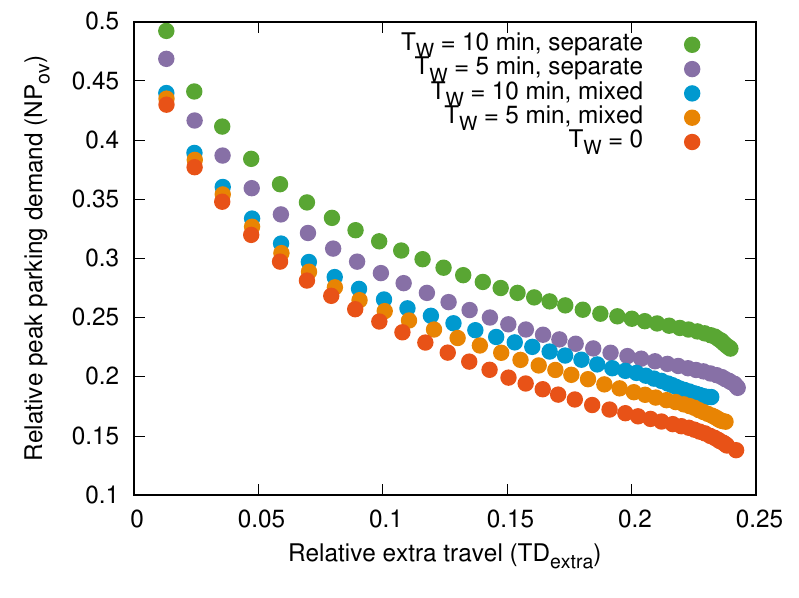}
		\caption{Estimating short-term parking requirements.
		Left: pick-up demand (i.e.~parking spaces used for short-term waiting) relative to the baseline parking demand as a function of the passenger convenience parameter ($T_W$). Absolute numbers reach up to 80,000 parking spaces for $T_W = 10\,\mathrm{min}$, but these correspond to larger and larger relative share. The curves start from $T_W = 1\,\mathrm{s}$, where the pick-up demand is the sum of the maximum number of trips starting at the same time at each discrete location, i.e.~about 8,000 parking spaces.
		Right: total parking demand ($NP_{ov}$) including pick-up demand as a function of increase in VKT ($TD_{extra}$) considering different expected extra required wait times ($T_W$) and usage policies for short-term parking. Note that these numbers also include the increased demand due to fleet size increases to compensate for the extra time spent idle. Results for $T_W = 0$ are the same as displayed in Fig.~\ref{newres1}.}
		\label{stparking}
	\end{figure*}

	We incorporate pick-up demand in our model via the passenger ``convenience'' parameter $T_W$ and require vehicles to arrive at least $T_W$ time before the start of each trip. Using this assumption, we calculate the number of short-term waiting spots needed to accomodate pick-up demand as the sum of the maximum number of waiting vehicles at each discrete location and interpret it as the ``price of convenience''. As displayed in Fig.~\ref{stparking} (left panel), total pick-up demand can be as large as 6\% of our current estimate of parking supply (for $T_W = 10\,\mathrm{min}$). Since pick-up demand is assumed to be at the exact location passengers start their trips, it is also highly affected by how such locations are discretized in space -- i.e., the extent to which pick-up locations are consolidated in a central point on each block or groups of blocks (see Fig.~\ref{stestimate} in the Supplementary Material).
		
	How much pick-up demand will add to the total parking demand of a city will depend on the policies related to use of short-term and long-term waiting areas, i.e.~whether pick-up areas can be used for long-term parking as well. We investigate two scenarios based on how pick-up areas are used. In Fig.~\ref{stparking} we show that if there is a strict separation between short-term and long-term parking (``separate case''), combined demand is significantly higher than if waiting areas are available for long-term parking as well (``mixed'' case).

	Our results show that a drastically reduced fleet of OVs could serve all private vehicle trips in Singapore, freeing up tremendous amount of space currently dedicated to parking. Parking lots could be redeveloped for housing and other productive uses, in turn stimulating the creation of dense, walkable neighborhoods and reducing urban sprawl~\cite{GonzalezGonzalez2019}. In the minds of some planners, OVs and AVs have the potential to spur an evolution towards a more pedestrian-oriented society and make urban living more attractive~\cite{NACTO2017,Shaver2019}. At the same time, reducing the size of the vehicle fleet has been linked to the potential of reducing lifecycle emissions from vehicle manufacture and disposal~\cite{Nealer2015,Greenblatt2015}.
		
	We do, however, identify four major limitations to the potential to reduce the amount of urban land devoted to parking. 
	
	First, the more urban land is freed up through parking consolidation, remote parking, and a shared fleet model, the more vehicle travel will result from deadheading and other vehicle relocation activity, e.g.~vehicles moving to and from the remote lots. While parking demand can be reduced to as little as 14\% of the baseline, this comes at the expense of a 24\% increase in VKT.
	
	Second, human mobility is temporally concentrated, particularly in the morning and afternoon peak commute periods~\cite{Alexander2015,Diao2016,Halvorsen2019}. Most vehicles will be idle at night, and many during the middle of the day, and require parking. While such parking facilities could be in remote locations, this would further increase vehicle travel from deadheading.

	Third, there is a spatial mismatch between the places where reduced parking is most beneficial, and the places where OVs can most efficiently contribute to parking reductions (see Figs.~\ref{spdist}--\ref{spdiff} in the Supplementary Material for the spatial distribution of parking). Most of the freed-up spaces are in residential areas, while the most valuable land is in the central business district.
	
	Fourth, pick-up demand can constitute a large amount of the total parking demand of an OV fleet, up to 25\% in highly optimized scenarios (see SI Fig.~\ref{stparking2}). As these locations need to be easily accessible to passengers, their placement and design is less flexible and can thus hamper the efficient parking solutions provided in the future by autonomous vehicles. At the same time, we expect that the size of pick-up areas can be limited if we assume on-demand vehicles to be identical and thus interchangeable. Passengers could start their trip in any available vehicle which are replaced by cars from long-term parking areas.
	
	In conclusion, our results show that very large swaths of urban land currently used for parking could be freed thanks to OD mobility. Furthermore, a range of trade-offs between parking reductions and increase in VKT have been quantified. While the equilibrium condition of a given city will depend on the actions of private fleet operators, governments have several tools to influence the latter’s behaviors. Deadheading can be discouraged by increasing the cost of vehicle travel through higher fuel taxes or congestion pricing. Parking can be discouraged in central areas through parking taxes and land-use planning; at a minimum, cities can eliminate distortionary regulations that require a minimum amount of parking~\cite{Shoup1995,Barter2011}, or establish maximums instead~\cite{ltaparking2}. Planning ahead using data-driven decision-making -- as enabled by approaches such as the one outlined in this paper -- is especially important when considering future scenarios where shared AV fleets might make OD mobility the predominant mode of transport~\cite{Krueger2016,Harper2016,Smith2012,Basu2018}.

\section*{Acknowledgments}
		
		The authors thank all sponsors and partners of the MIT Senseable City Laboratory including Allianz, the Amsterdam Institute for Advanced Metropolitan Solutions, the Fraunhofer Institute, Kuwait-MIT Center for Natural Resources and the Environment, 
		Singapore-MIT Alliance for Research and Technology (SMART) and all the members of the Consortium.
		
		This research is supported by the Singapore Ministry of Transport, Urban Redevelopment Authority, Land Transport Authority, Housing Development Board, Ministry of National Development and the National Research Foundation, Prime Minister’s Office under the Land and Liveability National Innovation Challenge (L2 NIC) Research Programme (L2 NIC Award No. L2NICTDF1-2016-3). Any opinions, findings, and conclusions or recommendations expressed in this material are those of the author(s) and do not reflect the views of the Singapore Ministry of Transport, Urban Redevelopment Authority, Land Transport Authority, Housing Development Board, Ministry of National Development and National Research Foundation, Prime Minister’s Office, Singapore.
		
\section*{Methods}
	
	\begin{figure*}
		\centering
		\includegraphics{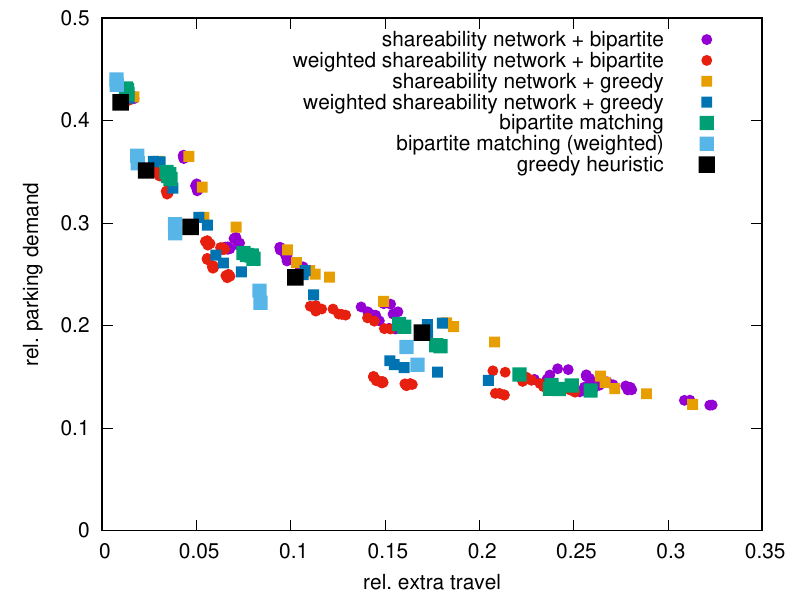}
		\quad
		\includegraphics{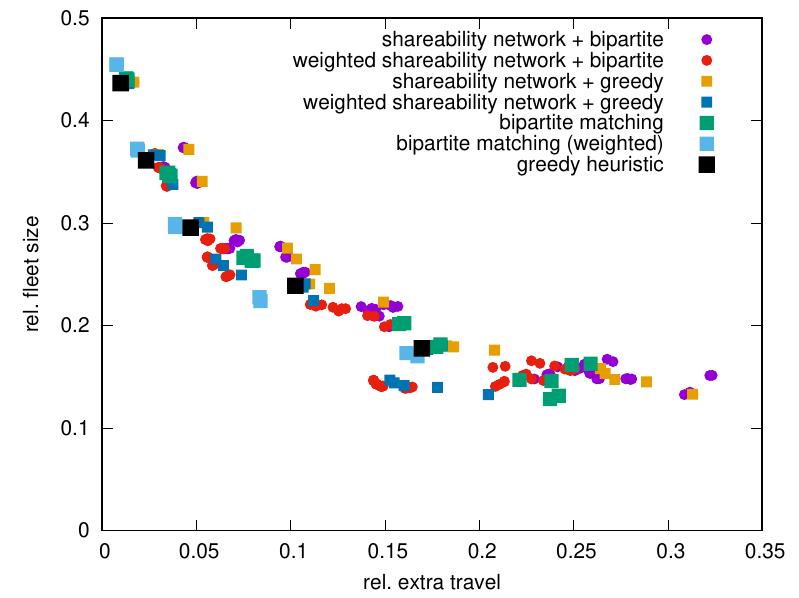}
		\caption{Comparison of results achieved with different approaches of assigning vehicles to trips and parking. Left: results for
			parking demand. Right: results for fleet sizes. For smaller fleet sizes, the approaches based on weighted matching tend to
			significantly outperform the other variations.}
		\label{res_all}
	\end{figure*}

	To estimate parking needs and travel of a fleet of OVs, we investigate the hypothetical scenario where every person currently using a private car for trips in the city-state of Singapore is willing to switch to a using a shared mobility service; we are thus exploring solutions which serve all current trips made by private vehicles. We estimate such trips using SimMobility, an integrated agent-based simulation platform for urban mobility capable of giving realistic estimates of trips made by the target population and calibrated to represent Singapore in 2012~\cite{Adnan2016}. Our methodology extends on our previous work focusing on a model of commuting~\cite{paper1} and methods employed by Santi~et~al.~\cite{Santi2014,Santi2018} with regards to ride-sharing and taxi fleet size estimation; now we focus on general trips for the whole population that are based on an extensive simulation of urban mobility in Singapore~\cite{Adnan2016}. Our main goal is to estimate the number of vehicles $NV_{ov}$, parking spaces $NP_{ov}$ and extra travel $TD_{extra}$ needed to accommodate all trips made in private vehicles under ideal conditions. We then compare our results to the current number of cars, parking and travel distance, denoted by $NV_{cur}$, $NP_{cur}$, and $TD_{cur}$, respectively. Note that $TD_{cur}$ is defined as the sum of the travel distance of the $NT$ given trips along shortest routes, under the assumption that there is always a parking spot available at the destination of a trip. Thus, $TD_{cur}$ can be considered as a baseline minimum distance that needs to be traveled to transport all passengers from respective origin to destinations. In this paper, we consider the variables representing OV system performance -- $NV_{ov}$, $NP_{ov}$, and $TD_{extra}$ -- as normalized vs.~the current system parameters. Formally:
\[
NV_{ov}\triangleq \frac{NV_{ov}}{NV_{cur}}, NP_{ov}\triangleq \frac{NP_{ov}}{NP_{cur}}, TD_{extra}\triangleq \frac{TD_{extra}}{TD_{cur}}~.
\]
	
	Our approach for fleet size estimation is in contrast to most previous work, which analyzed the operational characteristics of autonomous mobility on demand (AMOD) services under pre-determined fleet sizes~\cite{Fagnant2014,Fagnant2015b,Burns2013,Azevedo2016,Zhang2017}; our main interest is providing lower bounds on fleet size, parking requirements and travel and explore the tradeoffs among these bounds. In this goal, our work is most similar to that of Spieser~et~al.~\cite{Spieser2014} and Vazifeh~et~al.~\cite{Santi2018}. The main difference from Ref.~\cite{Spieser2014} is that our approach also results in an idealized dispatching strategy which satisfies all trips without delay, while the estimate presented in~\cite{Spieser2014} is an absolute minimum that does not take into account operational characteristics, resulting in a need for a much larger fleet to provide adequate service to most passengers. In contrast, the work of Vazifeh~et~al.~\cite{Santi2018} only focused on serving taxi trips and on ideal fleet size and did not consider parking or extra travel.
	
	Our focus on parking demand is most similar to the aims of the previous study by Zhang and Guhathakurta~\cite{Zhang2017}; the main difference again is that instead of running the simulation based on a presumed fleet size and parking availability, we aim to calculate the {\em minimum numbers} based on our constraints. Our work is complementary to recent work by Millard-Ball~\cite{MillardBall2019}, who present a detailed simulation of parking strategies on \emph{privately-owned} AVs and to Xu~et~al.~\cite{Xu2017} who investigate parking needs of ridesourcing vehicles. In contrast, we are explicitly interested in the scenario where the adoption of shared mobility brings along a shift in vehicle ownership as well.

	\subsection*{Trip data and estimate of current parking supply}
	
		The input for our analysis is a dataset of trips made by private vehicles generated by SimMobility, a complex platform for generating and simulating urban mobility realistically, based on a thorough process of calibration and verification using data including household travel surveys from the 2010-2015 period~\cite{Adnan2016}. Our dataset focuses on Singapore which is currently the main target of SimMobility. The data includes 1.44 million private car trips made by 676 thousand individuals over the course of one day in the simulation. This number is realistic for Singapore, a city-state of about 5 million people with one of the lowest number of private vehicles per capita in the developed world, but still suffering from the effects of congestion in peak periods and dedicating significant resources and space to road infrastructure. The trip data is generated by SimMobility's mid-term simulator module as trip chains taken by the agents in it, based on a calibrated model of present-day Singapore. We are not using SimMobility's capabilities to evaluate changes in mode share due the introduction of ride-hailing and AVs; instead, we are assuming that every trip made in private vehicles today would be substituted with a ride in an OV, thus we can investigate what are the implications of such a setup under simplified conditions.
		
		Besides trip data, SimMobility includes a database of buildings in Singapore~\cite{buildings,Le2016}. We estimate current parking supply from this database and publicly available data from official sources: the minimum parking requirements published by the Land Transport Authority (LTA)~\cite{ltaparking}, the list of parking spaces managed by the Urban Redevelopment Authority (URA)~\cite{uraparking}, and the aggregate number of parking spaces managed by the Housing Development Board (HDB)~\cite{hdbparking}. We further combine this data with results for parking occupancy from the trip data itself. This process results in an estimate of 1,369,576 parking spaces, or 2.03 parking spaces per person. We note that the real number is potentially even higher as our estimate is based on minimum requirements and the minimum amount needed to satisfy current trips made in private vehicles; in practice, developers may exceed the minimums. For comparison, parking supply for cities in the USA is estimated at between 2.49 -- 3.3 spaces per vehicle~\cite{Chester2015,Davis2010a,Davis2010b}. We note there are significant policy and land use differences between Singapore and the USA, thus it is reasonable that the Singapore levels are lower. Furthermore, as of 2019, Singapore has started significantly reducing minimum parking requirements and imposing maximum limits on parking for new development~\cite{ltaparking2}; nevertheless, since our simulations were calibrated based on data in the 2010-2015 period, it is reasonable that we use the prevailing minimum parking requirements at that time~\cite{ltaparking}. We describe the procedure for estimating the number of parking spaces in more detail in the Supplementary Material, in Section~4, with numbers of different types of parking displayed in Supplementary Table~\ref{tab_parking} and the spatial distribution of estimated parking supply in Fig.~\ref{spdist}.
	
	\subsection*{Trip estimation based on bipartite matching and heuristics}
		
		Our main methodology for assigning vehicles to trips and parking is given in detail in the Supplementary Material as Algorithms~1 and~2. It proceeds by first separating trips into start and end ``events'' and then processing these events sequentially (``greedy heuristic'', Algorithm~1) or in batches (``bipartite matching'', Algorithm~2). In both cases, an event is considered to be successfully processed if either (1) the end of a trip is matched to the start of a consecutive trip such that the same vehicle is able to serve the later trip after finishing the earlier one; (2) the start of a trip is matched to a parked vehicle, that is available to serve it; (3) the end of a trip is matched to an unoccupied parking location, meaning that the vehicle will park there after finishing the trip. Events that are not matched are then satisfied by adding more cars and parking to the simulation, similarly to previous works~\cite{paper1,Bauer2018}. This way, processing starts with zero cars and parking, creating only a minimal number of both over the course of processing all trips. This methodology also naturally allows us to keep track of vehicle utilization over the course of the simulation. We display the results of this in Fig.~\ref{fleet_utilization}.
		
		The main results of this paper were obtained by the bipartite matching methodology presented as Algorithm~2 in the Supplementary Material. Comparison of different methodologies is presented in Fig.~\ref{res_all}; results fit on the same trend as the main results displayed in Fig.~\ref{newres1}. A further variation is that each maximum matching problem can be performed as a weighted maximum matching by weighting the edges in the graphs by the distance of the connection. In this way, the result will minimize the connection distances while maximizing the number of matches. This results in general better results  (see Fig.~\ref{res_all}); we note that this comes at the cost of longer runtimes. We note that the greedy heuristic estimate (Algorithm~1) is an extension of our previous work~\cite{paper1}, while the main idea of performing a matching among trips in batches is similar to the online solution for taxi dispatching given by Vazifeh~et~al. in Ref.~\cite{Santi2018} and the online matching problem studied in detail by Lowalekar~et~al. in Ref.~\cite{Lowalekar2018}; the main difference and thus a necessary extension is to consider parking for vehicles not in use at any point in time.
		
		Beside matching trip requests in an online fashion or in batches, we also incorporated the global shareability network approach of Vazifeh~et~al.~\cite{Santi2018}. This corresponds to a model, where an ``oracle'' has knowledge of all trip demands in a day in advance and can decide on an optimal dispatching strategy based on that. We combined this approach with Algorithms~1 and~2 to be able to keep track of parking usage as well.
		
	\subsection*{Effect of discretization of space}
		
		The main results of the paper were obtained by considering a set of 4529 discrete ``nodes'' that can be the start and end locations of trips, with parking possibly present at any of these nodes. Travel distances and travel times were estimated between all node pairs based on the real road network and travel data. This raises the question if this discretization can affect the result of the simulation. We note that the discretization is present in a real city as well as there are a discrete set of buildings and associated parking locations and garages that can act as trip origins and destinations. Nevertheless, using only 4529 nodes is still an approximation. To test this, we also implemented a version of the main simulation in a continuous space model, where trips can start and end at any location, distances are taken as the Euclidean distance between points and travel times are calculated assuming a constant average travel speed. In this case, we adjusted the random start and end location of trips independently, by adding a variation to their coordinates chosen at uniformly random in an interval between $[-167\,\mathrm{m}, 167\,\mathrm{m}]$. We find that the results of this modified simulation agree well with the main results presented in this paper and increased fleet VKT can be modeled as an exponential function of the fleet size on an increased range of possible realizations. We display these results in the Supplementary Material in Figs.~\ref{res1} and~\ref{res_vcmp}.

\onecolumn
\newpage

\begin{center}
	{\Huge Addressing the ``minimum parking'' problem for on-demand mobility} \\[4ex]
	{\huge Supplementary Material} \\[3ex]
	{\large D\'aniel Kondor$^{1,\ast}$, Paolo Santi$^{2,3}$, Diem-Trinh Le$^{1}$,\\[0.5ex] Xiaohu Zhang$^{1,2}$, Adam Millard-Ball$^{4}$, Carlo Ratti$^{1,2}$ \\[1ex]
		\normalsize{$^1$Singapore-MIT Alliance for Research and Technology, Singapore}\\[0.3ex]
		\normalsize{$^2$Senseable City Laboratory, MIT, Cambridge MA 02139 USA}\\[0.3ex]
		\normalsize{$^3$Istituto di Informatica e Telematica del CNR, Pisa, Italy}\\[0.3ex]
		\normalsize{$^4$University of California Santa Cruz, Santa Cruz CA, USA}\\[0.3ex]
		\normalsize{$^\ast$ E-mail: \texttt{dkondor@mit.edu}}}
\end{center}

\renewcommand{\thefigure}{S\arabic{figure}}
\renewcommand{\thetable}{S\arabic{table}}

\setcounter{figure}{0}
\setcounter{table}{0}

\section{Main methodology}

In this section, we give a more detailed description of the methodology used to estimate fleet size and parking requirements for the on-demand vehicle service (OV) based on simulated trip data. The input of the methodology is a list of trips that people currently are assumed to make in private cars; we're assuming every trip will need to be served by OVs in the future. This way, the methodology examines a hypothetical scenario to characterize the maximum potential gains from the adoption of OVs. We note that the methodology does not depend on whether the OVs are autonomous or have a driver. Depending on which variation of the methodology is used, it relies either on the full knowledge of the trips in advance (i.e.~knowing all trips made in a day in advance) or at least partial advance knowledge, i.e.~knowledge of upcoming trip requests in a given time window (5-30 minutes, depending on the simulation parameters). While the latter case can correspond to a system requiring advance reservation, it is unclear whether such a system would be popular with users, who are now used to near-instantaneous reservations offered by transportation network companies (TNCs) and taxi reservation systems as well. Furthermore, all of our methodologies rely on a central controller to assign vehicles to passengers and parking.

\begin{figure}
	\centering
	\includegraphics{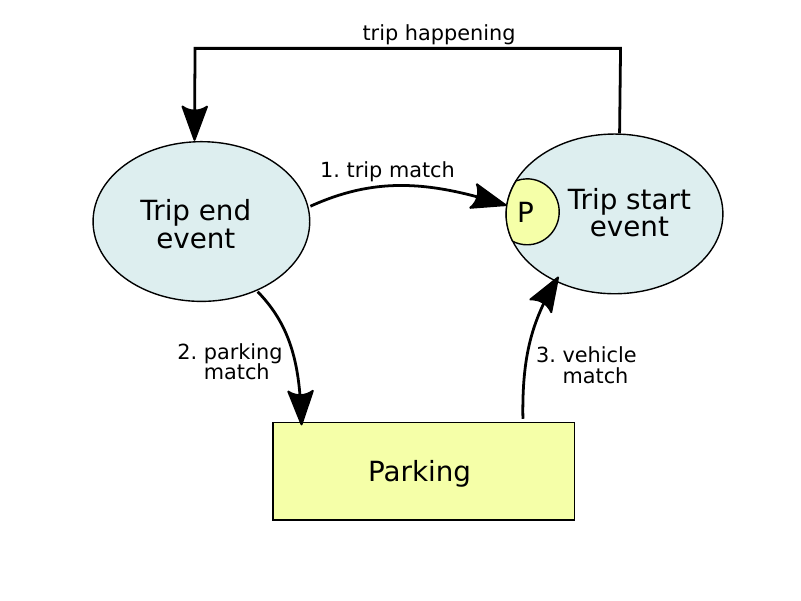}
	\caption{Left: illustration of the matching processes used to assign vehicles to trips and parking to vehicles.}
	\label{matching}
\end{figure}

Two variants of the main methodology for assigning vehicles to trips and parking spaces are outlined in Algorithms~\ref{alggreedy} and~\ref{algsteps} (notations used in both cases are further listed in Table~\ref{tabn} and further steps of Algorithm~\ref{algsteps} are detailed in Algorithm~\ref{algstepsmisc}). These can be further combined with the shareability networks approach to work on trip chains representing an ideal dispatching strategy instead of individual trips. Both variants work by assigning (1) trip ends to trip starts (i.e.~a vehicle makes a connection among consecutive trips); (2) trip starts to free cars (i.e.~vehicles are assigned to trips); (3) trip ends to free parking (free vehicles are assigned parking). This process is illustrated in Fig.~\ref{matching}. In both variants, when a trip start or end cannot be assigned a suitable free car or parking, a new car or parking space is ``created''. This way, the simulation starts with zero parking spaces and cars and only the minimum amount are added over the course of the simulation. The main difference between the two approaches is the way trips are processed: Algorithm~\ref{alggreedy} processes trips sequentially, always processing one trip start or end event at a time, assigning the closest available vehicle or parking (this way, we refer to it as a greedy heuristic). On the other hand, Algorithm~\ref{algsteps} processes trips in batches by using sliding time window that is progressed over the set of trips, performing optimized bipartite matchings in each step.

To make realistic decisions about vehicle assignment, we calculate the travel times based on real-world data as well. In the case of Singapore, average travel times between a set of road intersections were provided based on real traffic data at different times of the day and week. There are a total of 11,789 intersections, providing good coverage of the area.

\subsection{Preliminaries and notations}

Table~\ref{tabn} describes the main notations used in both algorithms. In both cases, we start with a set of trips ($T$), where each trip ($t \in T$) is a tuple defined by start location, end location, start timestamp and end timestamp. In this analysis, we represent locations by a set of discrete \emph{nodes}. Nodes can correspond to intersection, buildings, etc., and are defined by their coordinates and are matched to the road network. In the case of Singapore, we have a total of 4529 nodes. Travel times and distances (shortest path along the road network) among nodes are calculated in advance before running the algorithms described here. Our methodologies could easily accommodate more nodes or other representation of locations (e.g.~working directly with coordinates). Start and end timestamps of trips are considered fixed (and known in advance), thus discounting any potential variations in traffic. This also means that cars have to be available exactly at the start time of all trips. Each trip is separated into a start and end event, which are stored in two separate lists of events ($S$ and $E$ respectively). Both algorithms then work on these event lists. For a start or end event ($s \in S$ or $e \in E$), we then denote their timestamp as $t_s$ or $t_e$ and their location as $n_s$ or $n_e$. Since locations are represented as a discrete set of nodes, these can be directly compared for equality, while travel distances between locations are denoted by $d(e,s)$ and travel times by $t(e,s)$ and these are given by a lookup among the pre-computed values.

A main component of both algorithm are the list of available parking spaces and cars, denoted by $L_P$ and $L_C$ respectively. One parking space or car ($p \in L_P$ or $c \in L_C$) is represented by its location ($n_p$ or $n_c$, from the same set of nodes that were used by the trips) and a timestamp ($t_p$ or $t_c$) which is the time when the parking or car will become available. This is necessary since it will take some time for cars to reach parking or the start of the trips. E.g.~if a trip finishes at $t_e$ at node $n_e$, but a different node is selected for parking, the car will be only parked at time $t_c = t_e + t(e,c)$. When selecting the car to serve upcoming trips, this needs to be respected, i.e.~it can only serve trip for which $t_s > t_c + t(c,s)$. Note that, similarly to the notation used for the distances and travel times between trip start and end events, $t(c,s)$ and $t(e,p)$ denote travel time for a car to reach the start of a trip from its current location or a car to reach the parking stop after finishing a trip. Similarly, $d(c,s)$ and $d(e,p)$ denotes the travel distances involved.

\begin{table}[h]
	\centering
	\begin{tabular}{l|p{11cm}}
		notation & definition and usage \\ \hline
		$T$ & list of trips to be served (or trip chains previously constructed) \\
		$t \in T$ & one trip: each trip contains start and end locations and timestamps \\ \hline
		$S$ & list of trip starts\\
		$s \in S$ & one event in $S$, i.e.~the start of one trip, represented by location and timestamp; we require that a vehicle should be available at this location and time \\ \hline
		$E$ & list of trip ends \\
		$e \in E$ & one event in $E$, i.e.~the end of one trip, represented by location and timestamp; we require that the vehicle should be provided a parking location or assigned to a next trip \\ \hline
		$L_P$ & list of available parking spaces (initially empty) \\
		$p \in L_P$ & one available parking space; beside location, $p$ also contains a timestamp, which is the time when the parking becomes available (i.e.~this can be in the future for an already assigned trip) \\ \hline
		$L_C$ & list of available cars (initially empty) \\
		$c \in L_C$ & one available car; beside location, $c$ also contains a timestamp, which is the time when the car becomes available \\ \hline
		$t_x$ (for $x \in S,E,L_P\,\textrm{or}\,L_C$) & timestamp of entity (i.e.~start or end time of a trip, or timestamp when a parking or car becomes available) \\ \hline
		$n_x$ (for $x \in S,E,L_P\,\textrm{or}\,L_C$) & location of entity (start or end of trip or car or parking); locations are represented by one of discrete ``nodes'' \\ \hline
		$d(x,y)$ (for $x,y \in S,E,L_P\,\textrm{or}\,L_C$) & distance between the locations of two entities (e.g.~the distance from an available car to the start location of a trip) \\ \hline
		$t(x,y)$ (for $x,y \in S,E,L_P\,\textrm{or}\,L_C$) & travel time between the locations of two entities (e.g.~time it takes for an available car to reach the start location of a trip)
	\end{tabular}
	\caption{Notation used in Algorithms~\ref{alggreedy} and~\ref{algsteps}.}
	\label{tabn}
\end{table}

\subsection{Greedy heuristic estimation}

\begin{algorithm}
	\begin{algorithmic}[1]
		
		\State $T$ = \{ list of trips or trip chains \}
			
		\State $r_\textrm{max}$ = maximum distance that OVs are allowed to travel empty
		\State $t_\textrm{max}$ = look-ahead time for trip reservations

		\State $N_P = 0$ parking spaces required
		\State $N_C = 0$ number of cars required
		\State $D_{tot} = 0$ extra travel distance
		\State $L_P$ = \{ empty list for free parking spaces (along with timestamps when they become available) \}
		\State $L_C$ = \{ empty list for available cars (along with timestamps when they become available) \}
		
		\State $S$ = \{ empty event list for start events \}
		\State $E$ = \{ empty event list for end events \}
		\ForAll{$t \in T$}
				\State separate $t$ to start and end ``events''
				\State add these to $S$ and $E$ respectively
		\EndFor
		\State order $S$ and $E$ by timestamp
		
		\State process events in time order with $t_\textrm{max}$ shift between end and start events:
		\While{$S$ and $E$ not empty}
			\State $s$ = first element in $S$
			\State $e$ = first element in $E$
			\If{$t_s < t_e + t_\textrm{max}$}
				\Comment{process start event}
				\State $x = \arg\min_x d(x,s)$ where $x \in E$ s.t. $t_x \in \lbrack t_s-t_\textrm{max},t_s)$
					and $d(x,s) < r_\textrm{max}$ and
					\State \quad $s$ is reachable from $x$ (i.e.~ $t_x + t(x,s) < t_s$)
					\Comment find suitable matching end event
				 \If{found (i.e.~$x \neq \emptyset$)}
					\State remove $x$ from $E$
					\State reserve temporary parking at the location of $s$:
					\If{$\exists p \in L_P$, s.t. $n_p = n_s$ and $t_p < t_x + t(x,s)$}
						\State remove $p$ from $L_P$
						\State add parking to $L_P$ with $s$'s location and timestamp
					\Else
						\State $N_P = N_P + 1$
						\State add new parking to $L_P$ with $s$'s location and timestamp
					\EndIf
					\State $D_{tot} = D_{tot} + d(x,s)$ \Comment keep track of extra travel of the fleet
				 \Else \Comment{no matching end event, find closest available parked car}
					\State $c = \arg\min_c d(c,s)$ where $c \in L_C$ s.t. $d(c,s) < r_\textrm{max}$ and
						can reach $s$ (i.e.~$t_c + t(c,s) < t_s$)
					\If{found (i.e.~$c \neq \emptyset$)}
						\Comment assign $c$ to the trip
						\State remove $c$ from $L_C$
						\State add $c$'s location to $L_P$ (with the timestamp: $t_s - t(c,s)$)
						\State $D_{tot} = D_{tot} + d(c,s)$
					\Else
						\Comment add new car and parking to the system
						\State $N_C = N_C + 1$
						\State $N_P = N_P + 1$
						\State add new parking to $L_P$ with $s$'s location and timestamp
					\EndIf
				 \EndIf
				 \State remove $s$ from $S$
			\Else
				\Comment{process end event}
				\State $p = \arg\min_p d(e,p)$ where $p \in L_P$ s.t. $d(p,e) < r_\textrm{max}$ and is available (i.e.~$t_p < t_e + t(e,p)$)
				\If{found (i.e.~$p \neq \emptyset$)}
					\Comment assign $p$ as parking
					\State remove $p$ from $L_P$
					\State add $p$'s location to $L_C$ (with the timestamp $t_e + t(e,p)$)
					\State $D_{tot} = D_{tot} + d(e,p)$
				\Else
					\Comment add new parking to the system
					\State $N_P = N_P + 1$
					\State add car to $L_C$ with $e$'s location and timestamp
				\EndIf
				\State remove $e$ from $E$
			\EndIf
		\EndWhile
		
	\end{algorithmic}
	\caption{Greedy heuristic algorithm for determining fleet size and parking demand.}
	\label{alggreedy}
\end{algorithm}

Algorithm~\ref{alggreedy} describes a greedy heuristic to assign cars to trips and parking spaces. It is a simple and efficient way for achieving this and calculating the minimum number of cars and parking spaces required by starting from zero available and recording the ones that were added during the course of processing all trips. The basic functioning is to process trip start and end events in sequence and always assigns the closest available car to a trip start and the closest available parking space to a trip end. Travel time to and from parking spaces is taken into account, i.e.~it is recorded when a car reaches an assigned parking space and checked that it is able to reach the start of any new trip considered. A main parameter, $r_\textrm{max}$ determines the maximum distance cars are allowed to travel to reach trip starts or parking. E.g.~if $r_\textrm{max} = 1\,\mathrm{km}$ then at the start of the trip, a car is only assigned if the closest available car is less than $1\,\mathrm{km}$ distance away. If no car or parking is available closer than $r_\mathrm{max}$ when needed, a new one is added to the system, increasing the count of total number of cars or parking (this happens in lines~41-43 and~54-55 in Algorithm~\ref{alggreedy}). This way, we can start the simulation with zero cars and parking and only the minimum number will be added during the simulation.

A further consideration which is a separate part of Algorithm~\ref{alggreedy} is creating connections between consecutive trips. This means that instead of assigning a parking space after the end of a trip, a vehicle is assigned to serve an upcoming trip directly. This is an important part, since without this, a vehicle could be redirected to a parking space too far away, making it impossible to connect to an upcoming trip which happens closer or in a different direction. This is especially true for larger $r_\textrm{max}$ values, where travel to more distant locations is possible. Connecting trips is faciliated by introducing a ``look-ahead'' parameter, $t_\textrm{max}$, and assuming that reservations for upcoming trips are made at least this time in advance. The value for this is determined as $t_\textrm{max} = r_\textrm{max} / \bar{v}$, where $\bar{v}$ is a lower bound for the average expected travel speed assumed for short relocation trip. In practice, we use $\bar{v} = 20\,\mathrm{km} / \mathrm{h}$. Note that since we are assuming that trip start times are fixed (are served without delay), we are essentially requiring reservations in advance of $t_\textrm{max}$ even without trip connections, as this a bound on the time needed to reach a trip start from $r_\textrm{max}$ distance away. Analysing a system with live reservations, especially with respect to passenger waiting times is beyond the scope of the current work and has been investigated in several different contexts already.

To be able to process connections between trips, we process trip ends and starts separated by a window of $t_\textrm{max}$. This means that we process start events that are $t_\textrm{max}$ into the future compared to the end events being processed: selecting the next start and end events $s$ and $e$ from the time ordered lists $S$ and $E$, we select $s$ to process if $t_s < t_e + t_\textrm{max}$, and otherwise we select $e$ (this is indicated in line~20 of Algorithm~\ref{alggreedy}: processing of a trip start event happens in lines~21-46 and processing of a trip end event happens in lines~48-57). This way, when processing a trip start event, we not only look for cars which are parked, but also search among trip end events happening in the interval $[t_s - t_\textrm{max}, t_s)$. If there are any trip end events in this interval which are suitable (meaning that they happen less than $r_\textrm{max}$ distance away and the start event in question can be reached from them, i.e.~$t_e + t(e,s) < t_s$), we select the one with the shortest travel distance as the match and assign its vehicle to serve the trip start event in question (this is done in lines~24-33 in Algorithm~\ref{alggreedy}). In this case, the trip end event that was selected is not processed in the normal way to assign a parking to it (i.e.~it is removed from the event list, so lines~48-57 are not executed for it). Nevertheless, even in this case, temporary parking is reserved at the start of the trip as the vehicle will typically arrive earlier than the start of the trip it is assigned to. For this purpose, only parking available at the start location (start node) of the trip is considered. If no such parking is available, the total number of parking spaces is again increased.

The result of this analysis is the total number of parking spaces needed after processing all trips ($N_P$), the total number of cars needed (i.e.~fleet size, $N_C$) and the total ``extra'' distance traveled by the fleet ($D_{tot}$; this only includes distance traveled outside of making the trips themselves). Furthermore, the locations in $L_P$ and $L_C$ at the end of the simulation gives the locations where parking was needed during the simulation, thus our estimate of the spatial distribution of parking demand.

A further modification of Algorithm~\ref{alggreedy} is that it can be combined with an ``ideal'' dispatching strategy fonud by the methodology of \emph{shareability networks}. In this case, first a dispatching strategy is generated by creating a bipartite graph among trip ends and starts and calculating a maximum matching on it; the result of this is a list of trip chains, each chain being a sequence of trips that are served by the same vehicle. In this case, we use these trip chains as the input instead of individual trips. This means that start and end events ($S$ and $E$) will only include the start and end of whole trip chains instead of individual trips. In addition, since there could be need for temporary parking during a trip chain, we modify Algorithm~\ref{alggreedy} to process trip connections in the chain (supplied as separate input) in a similar manner to connections among trips found during running it, by reserving temporary parking at the start of trips in a chain (lines 25-33).

\subsection{Bipartite matching in batches}

\begin{algorithm}
	\begin{algorithmic}[1]
		
		\State $T$ = \{ list of trips or trip chains \}
			
		\State $r_\textrm{max}$ = maximum distance that OVs are allowed to travel empty
		\State $T_s$ = time window for trips to process at a time
		\State $T_w$ = time window for trips to consider when performing matching ($T_w \geq T_s$)

		\State $N_P = 0$ parking spaces required
		\State $N_C = 0$ number of cars required
		\State $D_{tot} = 0$ extra travel distance
		\State $L_P$ = \{ empty list for free parking spaces (along with timestamps when they become available) \}
		\State $L_C$ = \{ empty list for available cars (along with timestamps when they become available) \}
		
		\State $S$ = \{ empty event list for start events \}
		\State $E$ = \{ empty event list for end events \}
		\ForAll{$t \in T$}
				\State separate $t$ to start and end ``events''
				\State add these to $S$ and $E$ respectively
		\EndFor
		\State order $S$ and $E$ by timestamp
		
		\State $t = \min{t_s : s \in S}$ \Comment initialize first timestamp
		
		\While{$S$ and $E$ not empty}
			\State process trips in $\lbrack t, t + T_s )$, while considering trips in $\lbrack t, t + T_w )$
			\Step{{\bf 1.} match trip ends to starts}
				\State $G = \Call{CreateGraphStart}{E,S,t,t+t_w}$
				\State $M = \Call{BipartiteMatching}{G}$
				\ForAll{$(e,s) \in M$} \Comment{Process matched pairs ($s \in S$ and $e \in E$)}
					\If{$t_e < t + T_s$} \Comment{If end event is in main time window, remove events matching}
						\State remove $e$ from $E$
						\State remove $s$ from $S$
					\Else \Comment{exclude $e$ and $s$ from the current step as a candidte matching}
						\State mark $e$ and $s$ as reserved (will be used in next iteration again, but not in the following steps)
						\State remove $(e,s)$ from $M$
					\EndIf
				\EndFor
				\State note: matches in $M$ will be processed later
			\EndStep
			\Step{{\bf 2.} match trip starts to existing cars}
				\State $G = \Call{CreateGraphStart}{L_C,S,t,t+T_w}$
				\State $M_S = \Call{BipartiteMatching}{G}$
				\ForAll{$(c,s) \in M_S$} \Comment{Assign cars to trip starts ($c \in L_C$ and $s \in S$)}
					\If{$t_s < t + T_s$} \Comment{Only process start events in the smaller time window}
						\State remove $c$ from $L_C$
						\State add parking to $L_P$ with $c$'s location and timestamp $t_s - t(c,s)$
						\State remove $s$ from $S$
						\State $D_{tot} = D_{tot} + d(c,s)$
					\EndIf
				\EndFor
				\State $G = \Call{CreateGraphStart}{L_C,S,t,t+T_s}$ \Comment re-do the matching in the shorter time window
				\State $M_S = \Call{BipartiteMatching}{G}$ \Comment for the remaining start events
				\ForAll{$(c,s) \in M_S$} \Comment{($c \in L_C$ and $s \in S$)}
					\State remove $c$ from $L_C$
					\State add parking to $L_P$ with $c$'s location and timestamp $t_s - t(c,s)$
					\State remove $s$ from $S$
					\State $D_{tot} = D_{tot} + d(c,s)$
				\EndFor
			\EndStep
			\Step{{\bf 3.} process remaining start events (by adding more cars)}
				\ForAll{$s \in S$ s.t. $t_s < t + T_s$ and $s$ was not reserved in step \#1 above}
					\State $N_C = N_C + 1$
					\State $N_P = N_P + 1$
					\State add parking to $L_P$ with $s$'s location and timestamp
					\State remove $s$ from $S$
				\EndFor
			\EndStep
			
\algstore{algsteps}
	\end{algorithmic}
	
	\caption{Stepwise matching algorithm for estimating fleet size and parking demand}
	\label{algsteps}
\end{algorithm}

\begin{algorithm}
	\begin{algorithmic}
\algrestore{algsteps}
			\Step{{\bf 4.} match trip ends to existing parking}
				\State $G = \Call{CreateGraphEnd}{E,L_P,t,t+T_w}$
				\State $M_E = \Call{BipartiteMatching}{G}$
				\ForAll{$(e,p) \in M_E$} \Comment{Assign trip ends to parking ($e \in E$ and $p \in L_P$)}
					\If{$t_e < t + T_s$} \Comment{Only process start events in the smaller time window}
						\State remove $p$ from $L_P$
						\State add car to $L_C$ with $p$'s location and timestamp $t_e + t(e,p)$
						\State remove $e$ from $E$
						\State $D_{tot} = D_{tot} + d(e,p)$
					\EndIf
				\EndFor
				\State $G = \Call{CreateGraphEnd}{E,L_P,t,t+T_w}$ \Comment re-do the matching in the shorter time window
				\State $M_E = \Call{BipartiteMatching}{G}$ \Comment for the remaining end events
				\ForAll{$(e,p) \in M_E$} \Comment{($e \in E$ and $p \in L_P$)}
					\State remove $p$ from $L_P$
					\State add car to $L_C$ with $p$'s location and timestamp $t_e + t(e,p)$
					\State remove $e$ from $E$
					\State $D_{tot} = D_{tot} + d(e,p)$
				\EndFor
			\EndStep
			\Step{{\bf 5.} process remaining end events (by adding more parking)}
				\ForAll{$e \in E$ s.t. $t_e < t + T_s$ and $e$ was not reserved in step \#1 above}
					\State $N_P = N_P + 1$
					\State add car to $L_C$ with $e$'s location and timestamp
					\State remove $e$ from $E$
				\EndFor
			\EndStep
			\Step{{\bf 6.} process previously matched trips, provide temporary parking}
				\ForAll{$(e,s) \in M$}
					\State reserve parking at $s$'s location
					\If{$\exists p \in L_P$ s.t. $n_p = n_s$ and $t_p < t_e + t(e,s)$}
						\State remove $p$ from $L_P$
						\State add parking back to $L_P$ with $s$'s location and timestamp
					\Else
						\State $N_P = N_P + 1$
						\State add new parking to $L_P$ with $s$'s location and timestamp
					\EndIf{}
					\State $D_{tot} = D_{tot} + d(e,s)$
				\EndFor
			\EndStep
			\State $t = t + T_S$
		\EndWhile
		
	\end{algorithmic}
\end{algorithm}

In contrast to Algorithm~\ref{alggreedy} which performs local optimization by considering one trip at a time, we present an alternative approach a Algorithm~\ref{algsteps} which performs global optimization in batches, i.e.~considering trips that fall into time windows a given size. The main idea is that trips (both starts and ends) in a window of $T_w$ are considered and bipartite graphs are created from the possible connections (both among the trips themselves and among available cars and trip starts and trip ends and available parking). Then a maximum matching calculated on these bipartite graphs will result in a globally optimal solution for the assignment problem in that time window. Repeating the process with a sliding time window gives the solution for the whole set of trips considered.

In more detail, we create three bipartite graphs: (1) $G_1$, where edges connect trip end events to start events of later trips that can be reached from them; (2) $G_2$, where edges connect currently parked vehicles to upcoming trip start events that can be reached by them; (3) $G_3$, where edges connect trip end events to available parking spaces. Reachability is estimated using a dataset of estimated travel times between 11,789 intersections covering Singapore based on real traffic data including variation over a typical workday. The three graphs are created in three consecutive stages ($G_1$, $G_2$, $G_3$), each stage only including events that were not matched in the previous one: first, a matching is calculated among trip ends and starts in the time window under consideration (step~{\bf 1} in Algorithm~\ref{algsteps}, lines~21-32). The result of this are direct connections among trips that only require temporary parking (handling the requirement for these temporary parking is done in step~{\bf 6} in lines~85-95). Second, a matching among available cars and trip start events is performed (step~{\bf 2}, lines~34-51). Third, a matching among trip ends and available parking spaces is performed (step~{\bf 4}, lines~60-77). Trip start and end events which are not matched in any of the three stages are then satisfied by adding more cars and parking to the simulation, similarly to previous works~\cite{paper1,Bauer2018}. This way, processing starts with zero cars and parking, creating only a minimal number of both over the course of processing all trips.

A main parameter in these estimations is $r_\mathrm{max}$, the maximum distance cars are allowed to travel without a passenger (i.e.~``empty distance''). When creating each $G_i$ graph, only connections which include less than $r_\mathrm{max}$ travel are included; this way, we can limit the extra distance traveled by the fleet of OVs, although at the cost of more missed connections, thus requiring us to add more cars and parking. We repeat the processing with several different values of $r_\mathrm{max}$, ranging between $500\,\mathrm{m}$ and $10\,\mathrm{km}$.

Further parameters affecting the result are the batch time window $T_s$ and the look-out time window $T_w$. Among these, $T_s$ is the time window in which trips are processed at a time, while $T_w$ (with $T_w \geq T_s$) is the time window in which trips are considered. This means that matching is performed for all trips in the larger, $T_w$ time window to result in a more optimal solution, while the solution is only accepted and processed in the smaller, $T_s$ window. Trips which fall outside this smaller time window are processed in the next steps of the algorithm; after each step, time is advanced by $T_s$. A slight exception to this is matching trip start and end in steps~{\bf 2} and~{\bf 4}. In these cases, matching is performed twice: first for all events in $T_w$ with processing the solution for those that fall in the smaller window $T_s$. In the second step, the same matching is performed only for any trips remaining in the smaller $T_s$ window; the reason for this is to minimize adding more cars and parking spaces (i.e.~prioritize serving events in the present instead of setting aside an excess amount of resources for events in the future). In practice, we ran the algorithm with several options for $r_\textrm{max}$ between $500\,\mathrm{m}$ and $10\,\mathrm{km}$. We used $T_w = 15\,\mathrm{min}$, while we explored the choices of $5\,\mathrm{min}$ and $10\,\mathrm{min}$ for $T_s$.

Algorithm~\ref{algsteps} relies on several subroutines; among these, the functions for creating the bipartite graphs from a set of events are listed separately as Algorithm~\ref{algstepsmisc}; furthermore, standard solutions are used for calculating matching on bipartite graphs such as the Hopcroft-Karp algorithm. A potential further addition is calculating weighted bipartite matching (i.e.~minimum cost assignment problem) in each of steps~{\bf 1}, {\bf 2} and~{\bf 4}, using the connection distances as weights. This way, the solutions will attempt to minimize the extra travel taken by the fleet as well; we note that this version has significantly higher computational requirement, while resulting in somewhat better solutions.

\begin{algorithm}[h]
	\begin{algorithmic}[1]
		\Function{CreateGraphStart}{$X,Y,t_1,t_2$}
			\State $G = $ empty bipartite graph
			\ForAll{$x \in X$ s.t. $t_x \in \lbrack t_1, t_2 )$ and $x$ was not reserved in step \#1 of Algorithm~\ref{algsteps}}
				\ForAll{$y \in Y$ s.t. $t_y \in \lbrack t_1, t_2 )$ and $y$ was not reserved in step \#2 of Algorithm~\ref{algsteps}}
					\If{$t_x + t(x,y) < t_y$}
						\State add $(x,y)$ edge to $G$
					\EndIf
				\EndFor
			\EndFor
			\State \Return $G$
		\EndFunction
		
		\Function{CreateGraphEnd}{$X,Y,t_1,t_2$}
		\State $G = $ empty bipartite graph
			\ForAll{$x \in X$ s.t. $t_x \in \lbrack t_1, t_2 )$ and $x$ was not reserved in step \#1 of Algorithm~\ref{algsteps}}
				\ForAll{$y \in Y$ s.t. $t_y \in \lbrack t_1, t_2 )$ and $y$ was not reserved in step \#2 of Algorithm~\ref{algsteps}}
					\If{$t_x + t(x,y) > t_y$}
						\State add $(x,y)$ edge to $G$
					\EndIf
				\EndFor
			\EndFor
			\State \Return $G$
		\EndFunction
	\end{algorithmic}
	\caption{Functions to create the bipartite graphs on for matching in Algorithm~\ref{algsteps}. For the sake of a simpler presentation, the logic of these function is shown as a nested loop, having a complexity of $O(|X| \times |Y|)$. When implementing these, we use a more efficient solution which employs an index based on travel times, thus the actual complexity is $O(|G|)$ (i.e.~proportional to the number of edges returned).}
	\label{algstepsmisc}
\end{algorithm}

\begin{table}
	\centering
	\begin{tabular}{r|r|r|r}
		$r_\mathrm{max} [\mathrm{km}]$ & fleet size & parking requirements & ratio \\ \hline
		current situation	& 676,227	& 1,369,576	& $2.03$ \\ \hline	
		0.5			& 297,674	& 588,768	& $1.98$ \\
		1			& 234,384	& 476,511	& $2.03$ \\
		2			& 178,099	& 367,536	& $2.06$ \\
		5			& 121,196	& 247,621	& $2.04$ \\
		1			& 98,802	& 194,421	& $1.97$ \\
		unlimited	& 88,891	& 189,026	& $2.13$ \\
	\end{tabular}
	\caption{Results for fleet size, parking requirements and the ratio of the two for the bipartite matching methodology (main results in the main text, i.e.~Fig.~1).}
	\label{tab:ratios}
\end{table}

	\subsection{Discrete and continuous space simulation}
	
		In the results displayed in the main text (Figs.~1 and~6), the above calculations were performed using a set of 4529 discreted locations (``nodes'') as potential origins and destinations of trips. To better understand the relevance of this discretization, we also ran a modified simulation where each trip start and end event was assigned a random location independently in the neighborhood of the corresponding node (by adding a displacement to each coordinate, selected in a uniformly random way within $[-167\,\mathrm{m}, 167\,\mathrm{m}]$). Contrary to the main results, in this case, we assumed a simple Euclidean geometry, and a fixed average travel speed for OV operations that we varied between $10\,\mathrm{km}/\mathrm{h}$ and $50\,\mathrm{km}/\mathrm{h}$. We were mainly interested in extending the simulation to smaller values of the $r_\mathrm{max}$ parameter, where the discretization can become especially relevant. While $r_\mathrm{max} < 500\,\mathrm{m}$ is less relevant from an operational viewpoint, extending the analysis to this regime allows us to better characterize the dependence among the resulting values of fleet size and additional VKT.

\begin{figure*}
		\centering
		\includegraphics{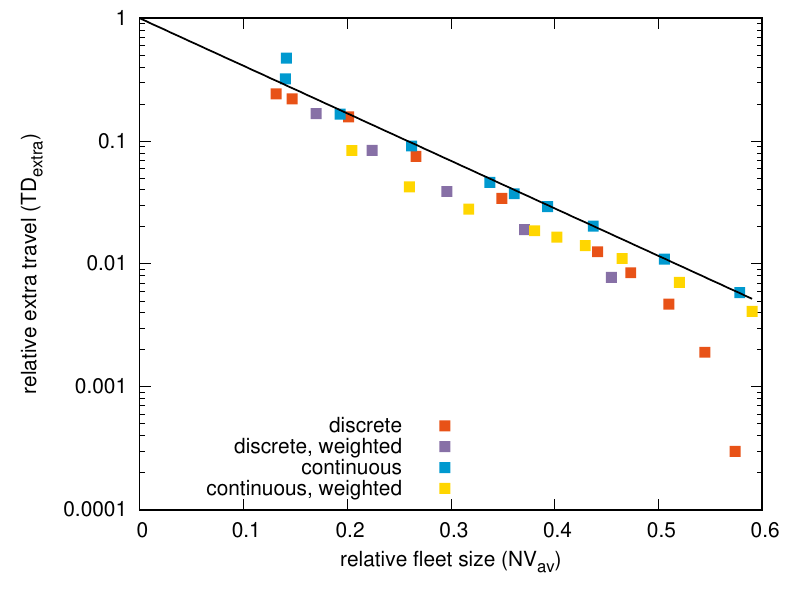} \,
		\includegraphics{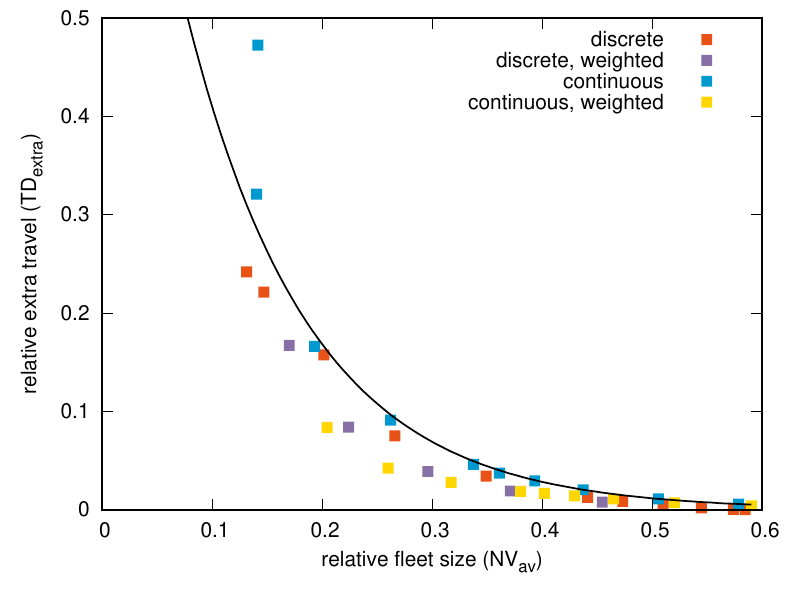}
		\caption{Total extra travel for the fleet as a function of fleet size for the discrete case (original simulations) and the simplified, continuous case (Euclidean space). Results are displayed for $r_\textrm{max}$ values between $50\,\mathrm{m}$ and $10\,\mathrm{km}$. The black line displays an exponential fit to the continuous case. The two panels display the same data with logarithmic and linear $y$-axis on the left and right respectively.}
		\label{res1}
	\end{figure*}

		In Fig.~\ref{res1}, we compare results of this modified simulation using continuous geometry with the original results presented in the main text (as Fig.~1 -- note that here the axes are reversed). We display the original simulation results as the ``discrete'' case (red points), along with the modified simulation results (for an assumed average speed of $50\,\mathrm{km}/\mathrm{h}$) as the ``continuous'' case (blue points). Furthermore, we display results with weighted matching in both cases as well (yellow and purple points). We also display an exponential fit to the results of the continuous simulation as black lines. This exponential form fits the results for the continuous space simulation quite well for the most of the range of results. It also fits the results of the discrete case nicely for $r_\mathrm{max} \geq 500\,\mathrm{m}$. We speculate that the deviation of the discrete case results for smaller $r_\mathrm{max}$ (larger fleet size and smaller extra VKT) is an effect of discretization: for small $r_\textrm{max}$ values, the only possible connection will be between trips ending and starting at the same node for an increasing share of cases.
				
\begin{figure*}
		\centering
		\includegraphics{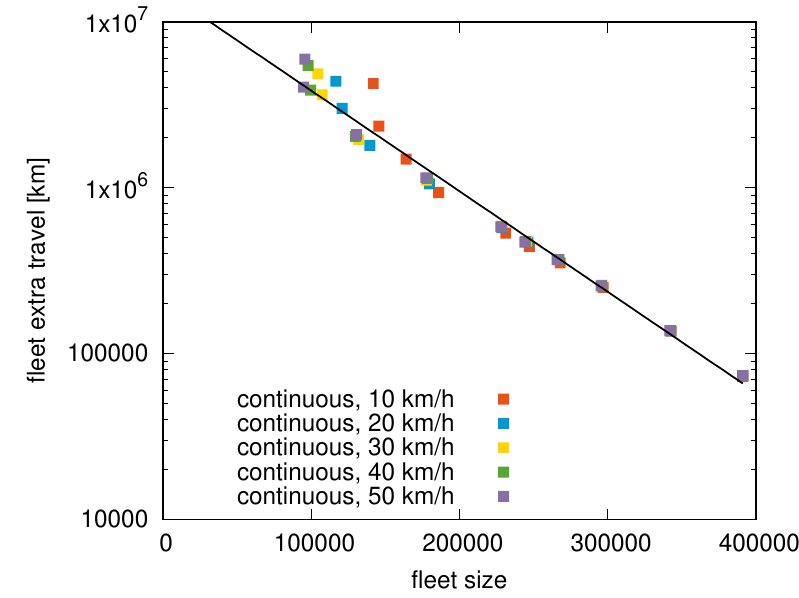}
		\caption{Total extra travel for the fleet as a function of fleet size for the continuous case, comparing results for different assumptions of average travel speed. We see that in this model, the average travel speed affects the results very little; there is a "saturation" effect limiting the minimum achievable fleet size as slower vehicle travel means less possible connections.}
		\label{res_vcmp}
	\end{figure*}

		In Fig.~\ref{res_vcmp}, we compare results for different values of the assumed average travel speed in the continuous space case. We see that these results are very similar, with a ``saturation'' effect present for small fleet sizes (large $r_\textrm{max}$).

	\subsection{Vehicle shareability networks}
		
		The vehicle shareability network provides a computationally efficient way to calculate an optimal dispatching strategy for a given list of trips, resulting in minimum fleet size or minimum total connection distances between trips~\cite{Santi2018}. In this approach, a network is constructed from the trips where each trip is a node, and two trips are connected by a (directed) edge if the start of the later trip can be reached from the end of the earlier one. This way, each edge represents a trip pair that can be served consecutively by the same vehicle. Each trip can have several incoming and outgoing edges as there can be several different options of which connections to choose at the end of a trip. Notably, trips separated by large time intervals can always be connected by an edge; this way, the shareability network can be huge, scaling with the square of the number of trips. We limit the network size and computational complexity by setting a maximum threshold on the connection time as well. Since we expect that a good dispatching strategy will prefer short connections, reasonable values of maximum connection time will only slightly affect the solution. Having constructed this shareability network, we can find an optimal dispatching strategy by finding a minimum path cover on it. Since the shareability network is a directed acyclic graph, finding a minimum path cover can be achieved efficiently (in polynomial time) by converting the problem into finding a maximum matching in a bipartite graph~\cite{Hopcroft, Boesch1977,Santi2018}. This minimum path cover can then be used as a dispatching strategy where each path is interpreted as a chain of consecutive trips to be served by the same vehicle. Thus, the number of vehicles needed is at most the number of paths found. We note that constructing the shareability network is equivalent to considering the $G_1$ graph in the previous method, but including all trips during the day (instead of only those in a limited time window $T_w$) and further limiting connection time instead of connection distance; this will result in some variation during the day due to variations in traffic speed.
		
		Having identified trip chains in an ideal dispatching strategy, we still need to assign parking to vehicles at the start and end of chains and inbetween consecutive trips. Furthermore, since demand fluctuates during the day, some trip chains can be distinct in time, thus possible to serve with the same vehicle (if separated by more than the minimum connection time). For these reasons, we continue by using the trip chains as the input of any of the previously described methods (``bipartite matching'' or ``greedy heuristic'', i.e.~Algorithm~2 or~1) to arrive at a final dispatching strategy that includes assignment of parking as well.
		
		While the result of this computation will be optimal in terms of fleet size, it can result in excessive extra travel as connection distance between consecutive trips in a chain is not part of the optimization. This problem is slightly mitigated by allowing only relatively short connection times, thus limiting connection distances as well. Furthermore, we can set an explicit limit on connection distances as well. An other approach that we implemented is again performing a weighted maximum matching after converting on the trip shareability network, giving a solution which minimized total connection distance as well.
		
		Looking at the results in Fig.~6 in the main text, we see that the results of this approach are mostly comparable with the previous ones; nevertheless, the combination of weighted shareability networks with the bipartite matching approach is able to significantly outperform all the other methods, resulting in 85\% reduction in fleet size and parking demand at a cost of only 15\% extra VKT as opposed to the results obtained by performing only unweighted bipartite matching, which achieves similar reductions at the cost of 25\% extra VKT. This outlines that performance in realistic conditions will be limited by operational constraints, i.e.~the fact that trip requests are generally not known in advance as required to construct shareability networks for the whole day and perform a global optimization.

\section{Incorporating ride sharing}

\begin{figure}
	\centering
	\includegraphics{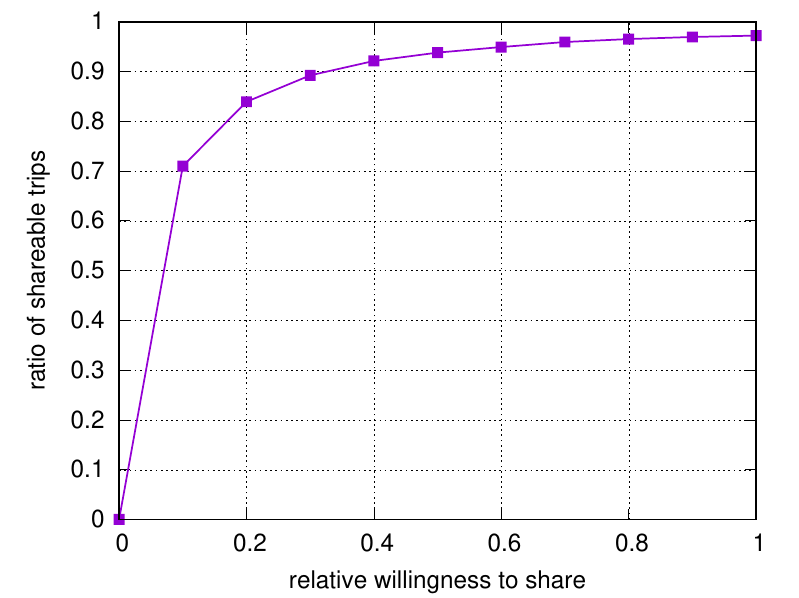}
	\caption{Shareability potential of trips. Ratio of trips actually shareable as a function of the ratio of trips where passengers would be willing to share their ride. Shareability ratio is relative to the trips available for sharing.}
	\label{shareability}
\end{figure}

\begin{figure}
	\begin{overpic}{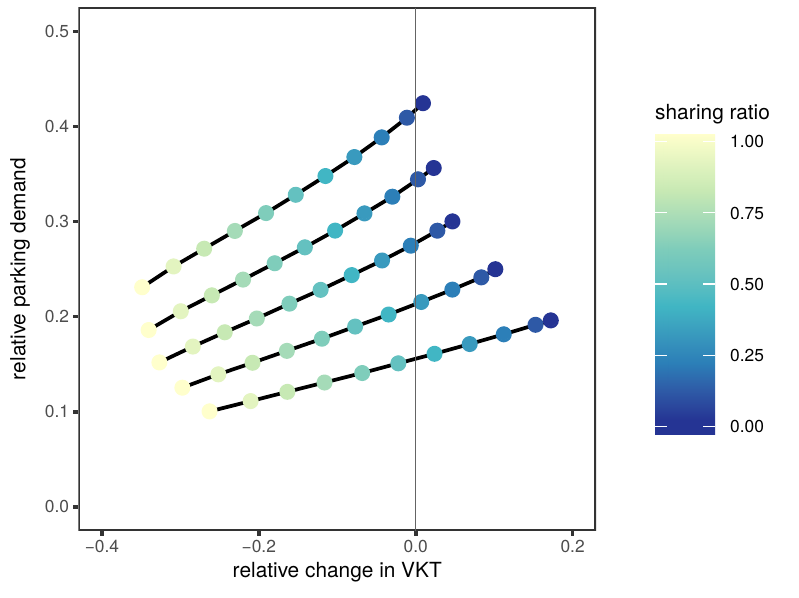}
		\put(54,64){\tiny $r_\mathrm{max} = 500\,\mathrm{m}$}
		\put(55,56){\tiny $r_\mathrm{max} = 1\,\mathrm{km}$}
		\put(59,48){\tiny $r_\mathrm{max} = 2\,\mathrm{km}$}
		\put(64,42){\tiny $r_\mathrm{max} = 5\,\mathrm{km}$}
		\put(60,28){\tiny $r_\mathrm{max} = 10\,\mathrm{km}$}
	\end{overpic}
	\begin{overpic}{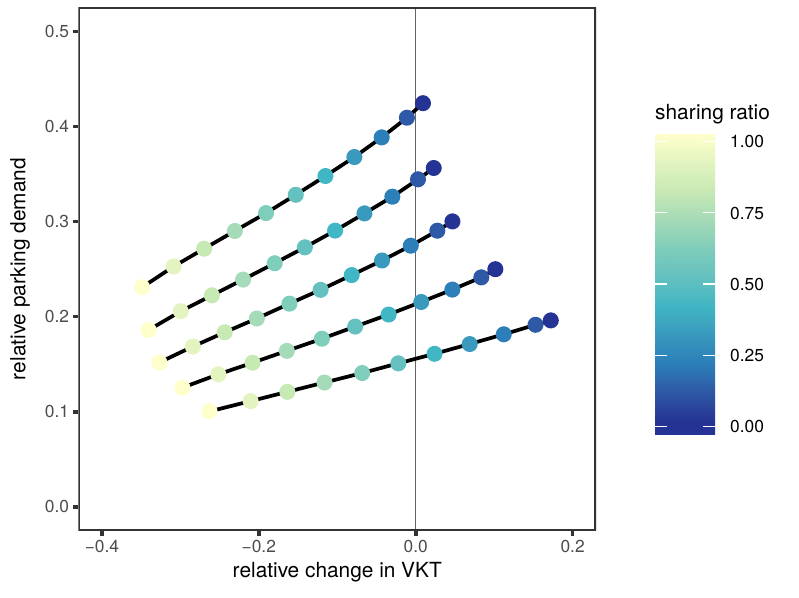}
		\put(54,64){\tiny $r_\mathrm{max} = 500\,\mathrm{m}$}
		\put(55,56){\tiny $r_\mathrm{max} = 1\,\mathrm{km}$}
		\put(59,48){\tiny $r_\mathrm{max} = 2\,\mathrm{km}$}
		\put(64,42){\tiny $r_\mathrm{max} = 5\,\mathrm{km}$}
		\put(60,28){\tiny $r_\mathrm{max} = 10\,\mathrm{km}$}
	\end{overpic}
	\caption{Results for fleet size, parking demand and change in total travel if people were willing to share rides. Color corresponds to the fraction of trips where people are willing to share rides (in 10\% increments from 0\% to 100\%); lines correspond to $r_\mathrm{max} = 500\,\mathrm{m}$, $1\,\mathrm{km}$, $2\,\mathrm{km}$, $5\,\mathrm{km}$ and $10\,\mathrm{km}$ from top to bottom respectively.}
	\label{ridesharing}
\end{figure}

A potential way to compensate for the extra travel of an AV fleet is to encourage people to share rides. We use the methodology of Santi~et~al.~\cite{Santi2014} to estimate the potential for shared rides under the assumption that $x$ percentage of people would be willing participate, with $x$ varying between 0\% and 100\%. Not surprisingly, we find that a large percentage of trips are shareable. We display the shareability ratio as a function of people willing to share (similarly to Ref.~\cite{Tachet2017}) in Fig.~\ref{shareability}. We see that if everyone was willing to share, over 97\% of all trips could be actually shared. Even if only 10\% of trips passengers are willing to share, the shareability ratio is still above 70\%.  For each case, we then produce merged trips that represent serving two original trips wherever sharing is possible. We then repeat our analysis for this set of combined trips (i.e.~merged trips and original trips that cannot be shared or passengers are unwilling to share) and display the results as a function of $x$ and $r_\textrm{max}$ in Figure~\ref{ridesharing}. We see that as $r_\textrm{max}$ gets larger (and dispatching gets more efficient in general), a larger $x$ share of shared rides are required to compensate for added VKT of a shared fleet.

\section{Estimating short-term parking requirements}
	
	Beside a change in the total number of parking, a further important consideration is a change in the use of parking on a local level. If people are willing to change from using private cars to OVs, actual parking facilities will be less important from the users' point of view (i.e.~they will not need to be easily accessible for average users), while the provision and design of adequate pick-up and drop-off locations will become increasingly important. Furthermore, in a future scenario with autonomous vehicles (AVs) we expect users' expectations to change regarding the start of their trips: since the cost of waiting can decrease significantly as it no longer includes a driver's salary, users might prefer to order a vehicle in advance and have the vehicle wait for them instead of having to wait themselves. While this increases convenience for passengers, booking trips some time in advance can be beneficial for operators as well, since this will allow the use of a dispatching algorithm that aggregates requests in time, allowing more efficient solutions. This way, vehicles waiting for passengers in pick-up locations could add a non-negligible amount of demand for temporary parking spaces. In the following, we consider that vehicles have to arrive exactly $T_W$ time before the start of each trip. We vary the $T_W$ parameter between $1\,\mathrm{s}$ and $10\,\mathrm{min}$.
	
	\subsection{Absolute demand for short-term parking}
		
		\begin{figure}
			\includegraphics[width=3.2in]{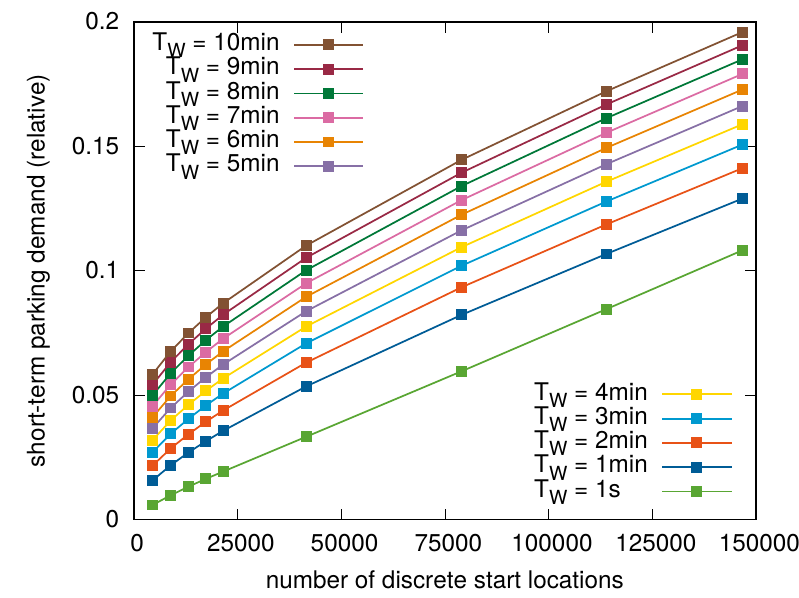}
			\includegraphics[width=3.2in]{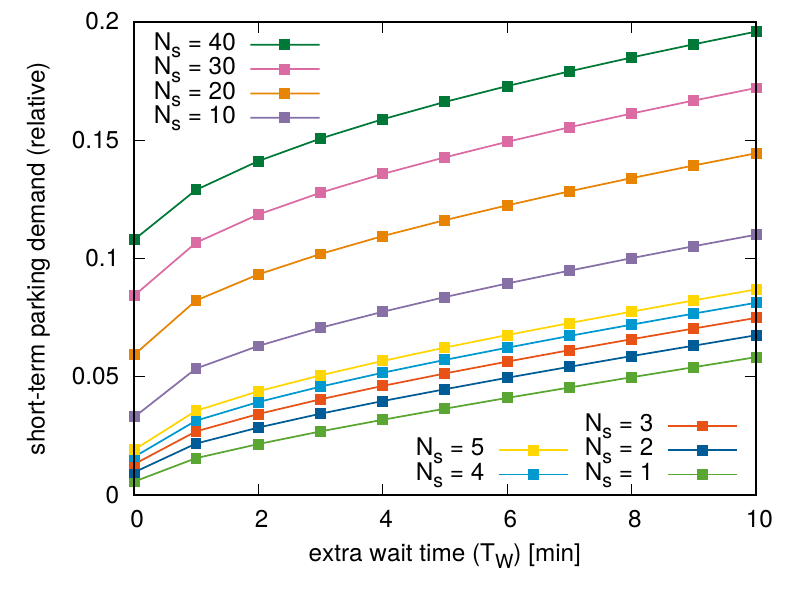}
			\caption{Estimation of pick-up demand (fraction of baseline parking demand). Left: pick-up demand for different choices of $T_W$ as a function of the number of discrete nodes considered as possible trip start locations. Right: same results displayed as the function of $T_W$ with distinct curves corresponding to different $N_s$ parameter values and thus different number of distince nodes. Values displayed in this figure are the average of 100 realizations of the random process described in the text. Standard deviation of values among the 100 realizations is less than 0.25\% in all cases.}
			\label{stestimate}
		\end{figure}
		
		\begin{figure}
			\centering
			\includegraphics{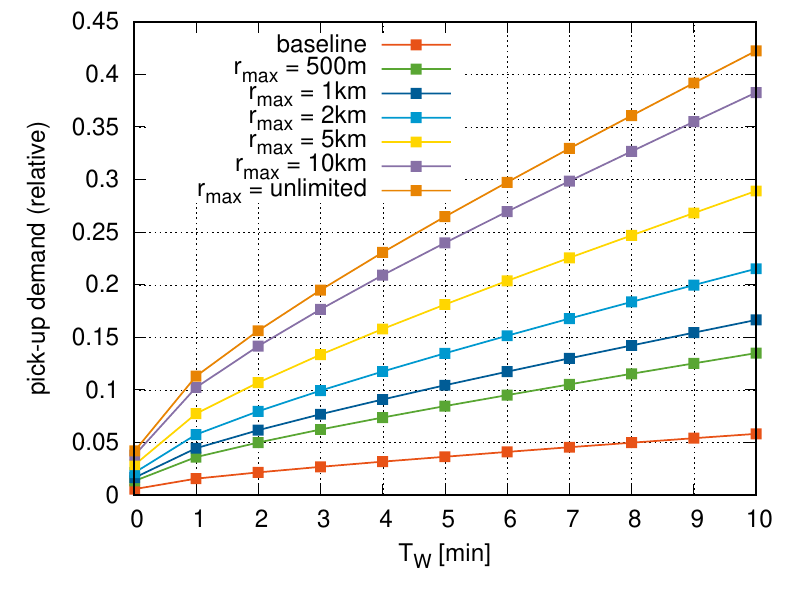}
			\caption{Pick-up demand (i.e.~parking spaces used for short-term waiting) relative to the parking demand estimated previously as a function of the assumed extra wait time ($T_W$). Absolute numbers are the same for each curve, reaching about 80,000 parking spaces for $T_W = 10\,\mathrm{min}$, but these correspond to larger and larger relative share. The curves start from $T_W = 1\,\mathrm{s}$, where the pick-up demand is the sum of the maximum number of trips starting at the same time at each discrete location, i.e.~about 8,000 parking spaces.}
			\label{stestimate12}
		\end{figure}
		
		Formally, for a trip $t$ that start at node $n_s$ at time $t_s$, we require the car serving it to be parked at $n_s$ in the interval $[t_s-T_W; t_s]$. Based only on this consideration, we can calculate the number of waiting cars at each node as
		\[
			N_W (n,t) \equiv \sum_s \delta_1 \left (\frac{t-t_s+T_W}{T_W} \right) \delta_{n_s,n}
		\]
		where the summation in $s$ includes all trip start events (refer to Table~\ref{tabn} for notations), $\delta_1$ is a $1$-width Dirac-delta function, while $\delta_{n_s,n}$ is the usual Kronecker delta symbol, defined as:
		\[
			\delta_1 (x) = \left \{ \begin{array}{ll}
				1 & \textrm{if } x \in [0,1] \\
				0 & \textrm{otherwise} 
				\end{array} \right .
			\quad \quad
			\delta_{x,y} = \left \{ \begin{array}{ll}
				1 & \textrm{if } x = y \\
				0 & \textrm{otherwise} 
				\end{array} \right .
		\]
		Using these, we can formally calculate the total demand for short-term parking as
		\[
			N_W^{\mathrm{total}} = \sum_{n \in \mathrm{nodes}} \max_t N_W(n,t)
		\]
		where the summation is done over all discrete nodes in the simulation. The result of this calculation is displayed in Fig.~5 (left panel) in the main text.
		
		Since we calculate the maximum number of waiting cars per nodes (i.e.~discrete locations), the discretization of space can affect the results for $N_W^{\mathrm{total}}$. Clearly, if all trips started in distinct locations, the number of parking places for waiting cars would equal to the number of trips (1.44 million in our case). Having multiple trips start at the same discrete locations (nodes) allows them to share the same parking space if their waiting time does not overlap. This way, using only 4529 discrete nodes as trip start and end locations can underestimate pick-up demand. To overcome this limitation, we repeated the previous estimation after splitting nodes in $N_s$ virtual locations and assigning trips to start at any of these at uniformly random. In Fig.~\ref{stestimate} we display pick-up demand ($N_W^{\mathrm{total}}$) estimated in these scenarios as a function of the total number of discrete nodes with at least one trip starting there (we note that the number of actual nodes is not necessarily $N_s \times 4529$ since some nodes have only a few trips starting there, so they do not generate $N_s$ separate nodes). While it is unclear how many discrete pick-up locations will be needed in the future, a reasonable upper bound is the total number of buildings in Singapore. Exploiting the fact that in Singapore, every building is assigned a unique postal code, we can estimate the total number of buildings based on a database of postal codes. We downloaded the list of unique post codes from the API of OneMap.sg~\cite{postcodesonemap} using the scripts published at~\cite{postcodesgit}, finding a total of 121,363 unique postal codes. This way, we estimate that the upper end of the results presented in Fig.~\ref{stestimate} show a realistic upper bound on the pick-up demand.
	
	\subsection{Combined estimate of total parking demand}
	
		While our previous analysis gives an estimate of pick-up demand by itself, we are also interested what is the combined effect on parking demand in our estimate based on trips. To do this estimation, we run a modified version of the main estimation (Algorithm~\ref{algsteps}) that takes into account the requirement for vehicles to wait for passengers before the start of trips. We evaluate two distinct scenarios regarding the use of short-term parking spaces:
		
	\begin{enumerate}
		\item Each location has a separate area reserved for short-term parking (i.e.~pick-up locations) beside a normal facility for long-term parking (e.g.~a parking garage). All vehicles that are waiting to pick up a passenger are required to park in the pick-up area, while vehicles are not allowed to park there long-term. The size of this area is determined over the course of the simulation, i.e.~we do not constrain it, but record the highest number of vehicles waiting at the pick-up area of each node at any time.
		\item Instead of reserving a short-term parking area only for vehicles waiting for a passenger, this area can be used for long-term parking as well. We further assume that all vehicles are essentially interchangeable, thus passengers need not be assigned any  specific vehicle, but will be able to start their trip in any of the vehicles available in the pick-up area. This way, we do not  have to explicitly divide parking into short-term and long-term facilities during the course of the simulation, but assume that vehicles always occupy the waiting area first and only park in the long-term facilities if the waiting area is full. For each discrete location, we record the maximum number of cars simultaneously waiting for passengers as the estimated demand for short-term parking. In this case, we can expect more optimal solutions in terms of combined (short-term and long-term) parking demand.
	\end{enumerate}

\begin{figure}
	\centering
	\vspace{2ex}
	\begin{overpic}{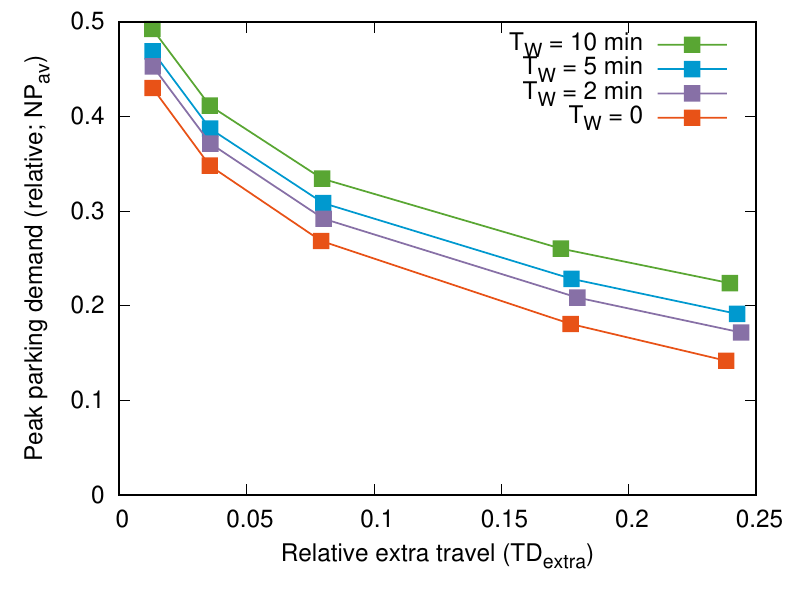}
		\put(55,77){\makebox(0,0){\sffamily Separate short-term parking (scenario \#1)}}
	\end{overpic} \,
	\begin{overpic}{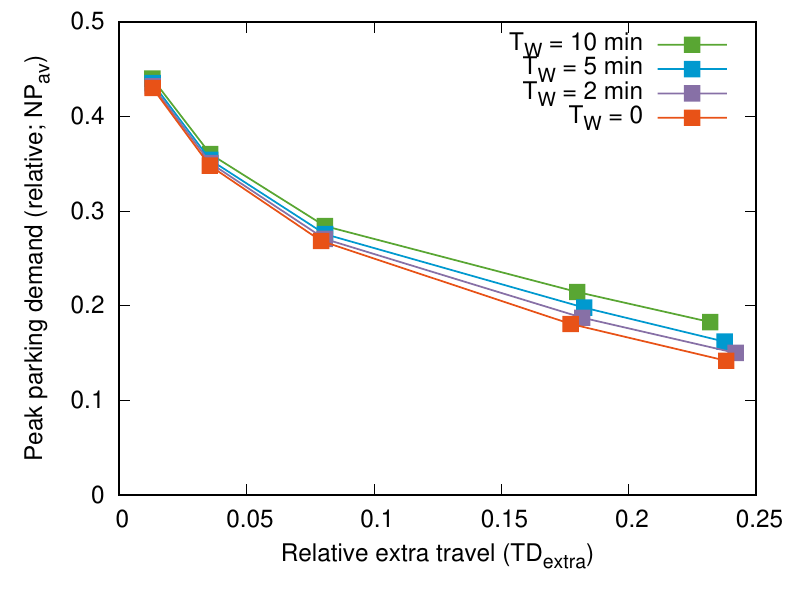}
		\put(55,77){\makebox(0,0){\sffamily Mixed parking (scenario \#2)}}
	\end{overpic}
	\caption{Estimating parking requirements including pick-up demand. Left: scenario \#1, i.e.~short-term and long-term parking are stricly separated. Right: scenario \#2, i.e.~pick-up areas can be used for long-term parking as well.}
	\label{stparking1}
\end{figure}

\begin{figure}
	\centering
	\vspace{2ex}
	\begin{overpic}{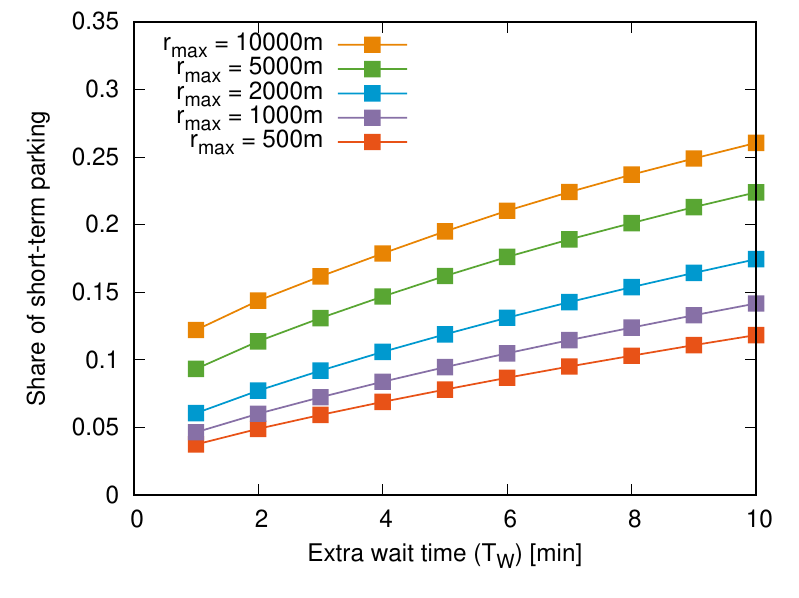}
		\put(55,77){\makebox(0,0){\sffamily Separate short-term parking (scenario \#1)}}
	\end{overpic} \,
	\begin{overpic}{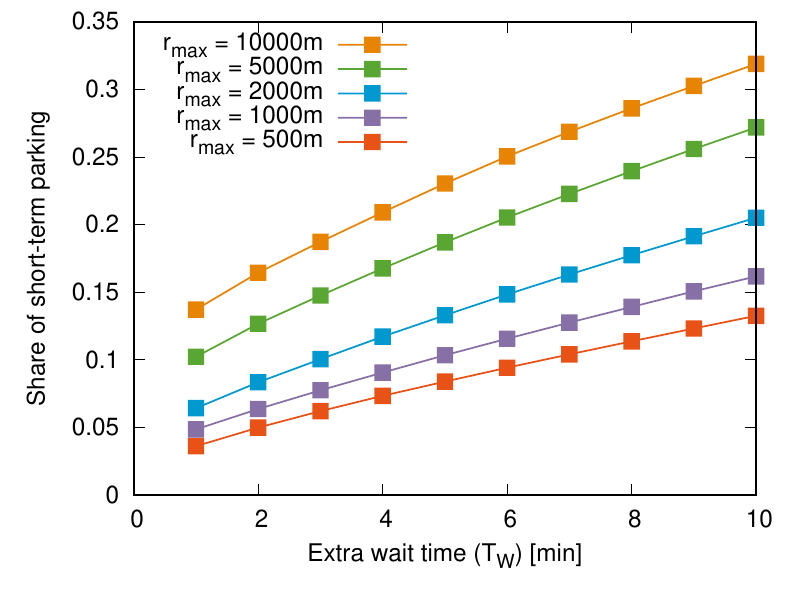}
		\put(55,77){\makebox(0,0){\sffamily Mixed parking (scenario \#2)}}
	\end{overpic}
	\caption{Relative share of pick-up demand (parking spaces used at least once for short-term wait) among the \emph{combined} parking demand in both scenarios as a function of the extra wait time ($T_W$). Contrary to Fig.~5 in the main text where the pick-up demand was compared to the long-term parking demand without considering short-term waiting at all, here we display the share of parking spaces used for short-term waiting after running the modified simulation that takes into account extra wait as well.}
	\label{stparking2}
\end{figure}
	
	We note that the actual pick-up demand (parking spaces needed to accommodate waiting vehicles) will be the same in both scenarios, as the number of vehicles waiting at each node is only determined by the sequence of trips starting there and the $T_W$ parameter, as estimated before. In the first scenario, this is added to the long-term parking demand, while in the second scenario, there can be an overlap, thus the total increase will be potentially less than the total need for short-term parking. In both cases, there is a potential increase in fleet size since vehicles need to spend some of their time waiting; in turn, this can add to long-term parking requirements as well.
	
\begin{figure}
	\centering
	\vspace{2ex}
	\begin{overpic}{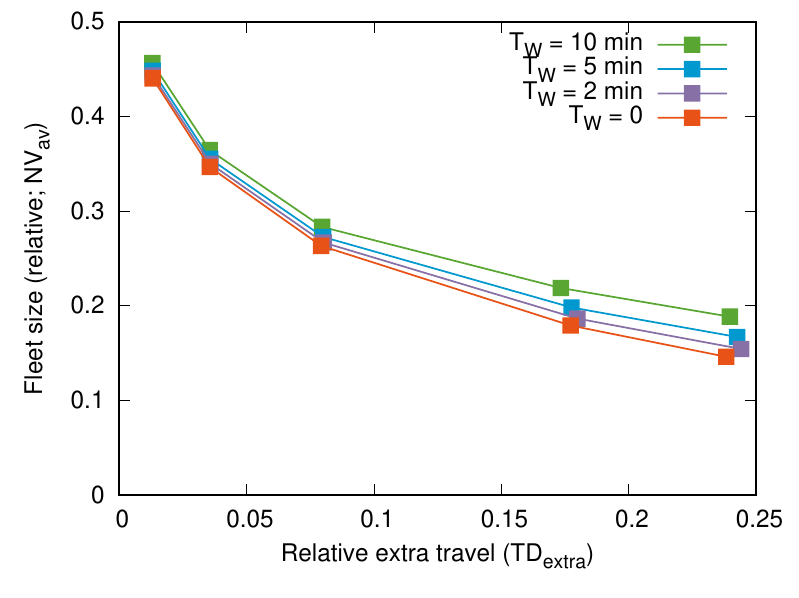}
		\put(55,77){\makebox(0,0){\sffamily Separate short-term parking (scenario \#1)}}
	\end{overpic} \,
	\begin{overpic}{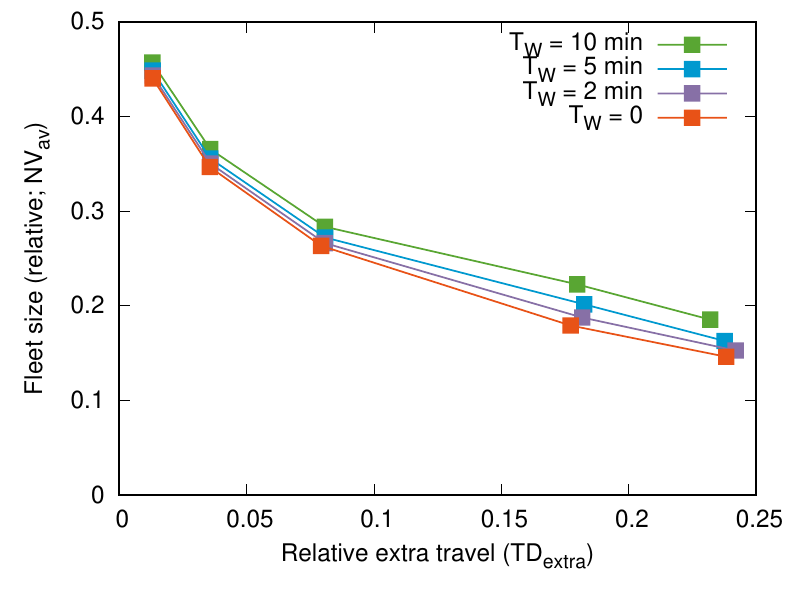}
		\put(55,77){\makebox(0,0){\sffamily Mixed parking (scenario \#2)}}
	\end{overpic}
	\caption{Change in fleet size requirement due to extra wait before the start of trips.}
	\label{stparking3}
\end{figure}

	Again, we vary $T_W$ between $0$ and $10$ minutes, with $T_W = 0$ representing the original analysis without pick-up demand. Here in Fig.~\ref{stparking} we present the comparison among the two scenarios, while in Fig.~\ref{stparking2}, we show the relative share of short-term parking in scenario and the absolute number of short-term parking spaces. In Fig.~\ref{stparking3}, we display the change in fleet size in these cases.
	
	Not surprisingly, scenario~\#1 results in higher increases in total parking needs as there parking designated as short-term cannot be reused for long-term parking purposes. In either case, our results emphasize the importance of designing pick-up and short-term waiting areas in an efficient way.

	\subsection{Effect of short-term parking policy of parking ratio}
	
	As illustrated in Fig.~\ref{matching}, the processes used to assign vehicles to trips and parking (Algorithm~2) has three steps where maximum matching is performed to optimize utilization of vehicles and parking. First, the end of trips are matched directly to start of trips (\#1); this step will ensure maximum utilization for existing vehicles. Since a vehicle assigned to a new trip might arrive too early, parking is always reserved at the start node of the new trip, even if we are not counting short-term parking separately. After this step, the remaining trip end and start events are matched to available parking and idle vehicles and  respectively (\#2 and \#3). Remaining trip end and start events are then satisfied by creating new vehicles and parking. Including the first step (trip matching) can help in finding more optimal solutions, since after finishing a trip, vehicles need not travel to parking in potentially distant locations, but can serve any new trip request close by.

\begin{figure}
	\centering
	\includegraphics{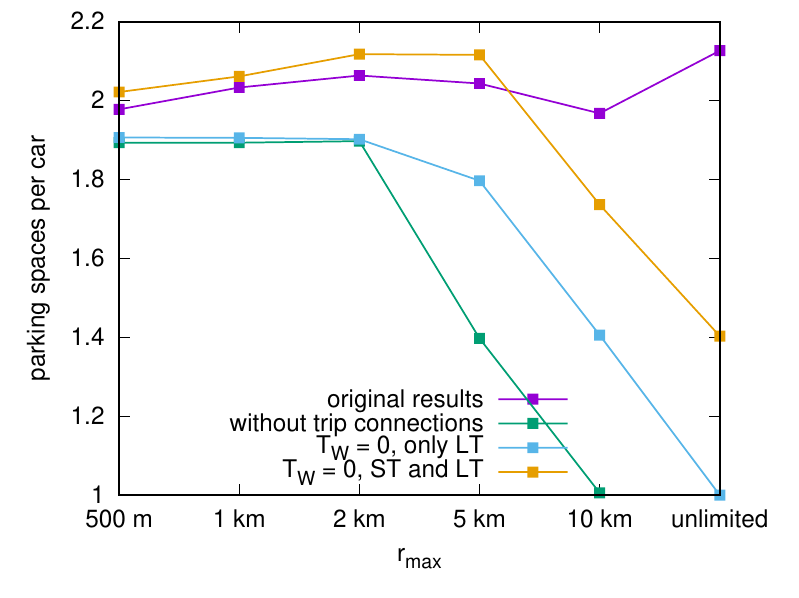} \,
	\includegraphics{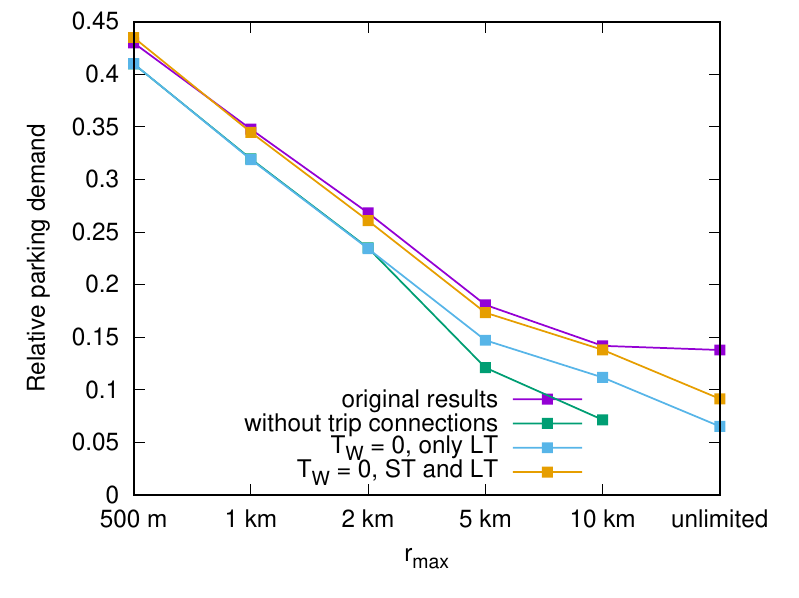} \\
	\includegraphics{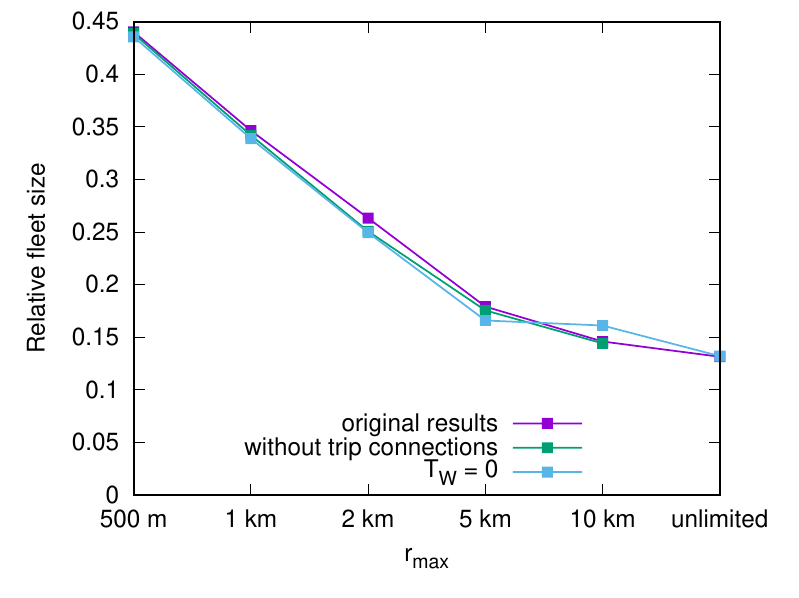} \,
	\includegraphics{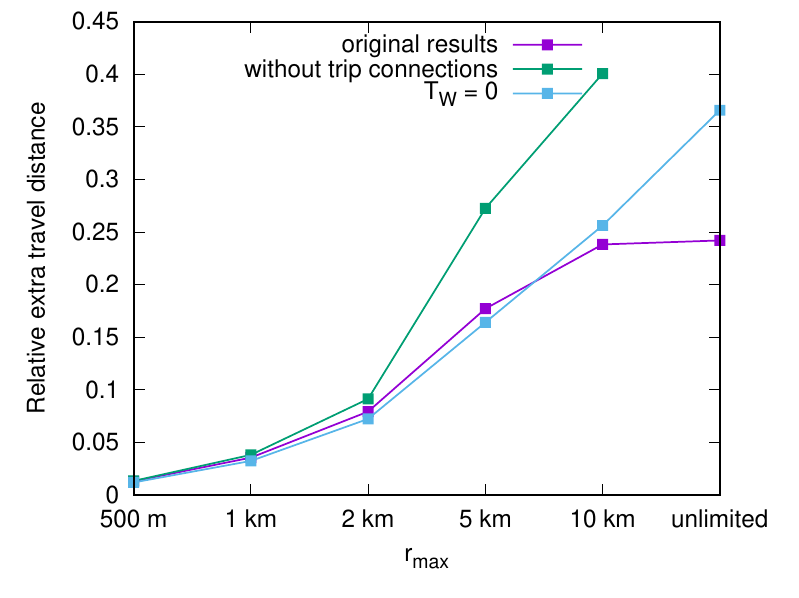}
	\caption{Comparing different methods to estimate demand for parking and fleet size. Ratio of parking to cars (top left), relative number of parking spaces (top right), relative fleet size (bottom left), and relative extra VKT (bottom right) are displayed as a function of $r_\mathrm{max}$. Among the cases presented here, ``original results'' refers to the bipartite graph approach presented as the main results in the paper (Fig.~1 and ``bipartite matching'' in Fig.~4); ``without trip connections'' refers to the same methodology, but omitting direct matching between consecutive trips (cars always have to return to parking at the end of a trip, new trips can only be matched to cars that are already parked); the ``$T_W = 0$'' cases refer to the estimation that has a strict separation of short-term parking, but with zero extra wait time; in this case, short term parking is only used by cars that are directly matched between the end and start of consecutive trips. In this case, ``only LT'' includes the count of only long-term parking demand, while ``ST and LT'' includes the sum of long-term and short-term parking demand.}
	\label{ratios1}
\end{figure}

\begin{figure}
	\centering
	\includegraphics{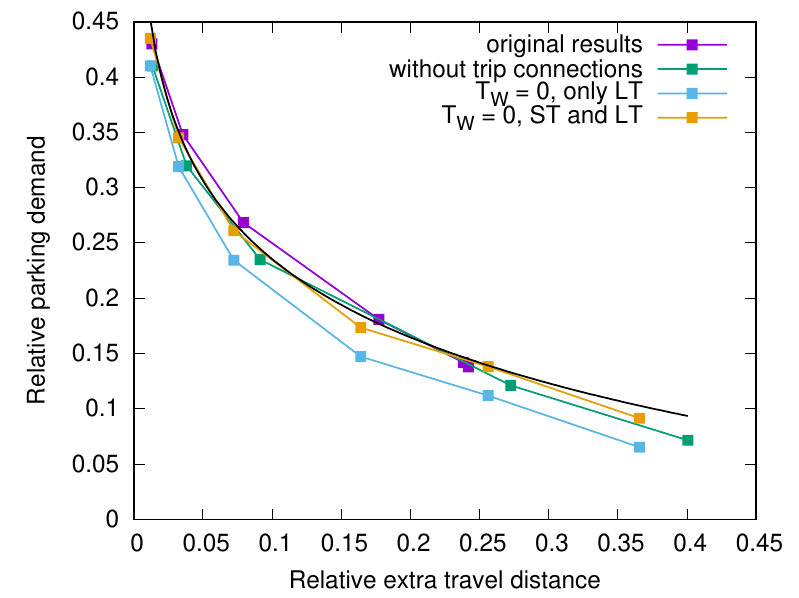}
	\caption{Relative parking demand as a function of extra VKT calculated with different approaches. We see that all approaches show very similar results that are consistent with the exponential fit in the main text (black line); the main difference is the case when we disregard short-term parking (``$T_W = 0$, only LT''), which is a clear underestimation of total parking requirements.}
	\label{ratios2}
\end{figure}

	We explored multiple options regarding how parking is accounted for. In the original estimation, parking is not differentiated between the first process (trip matching, which results in essentially short-term parking demand) and the second and third one (which matched trips to long-term parking). Thus, the first process will result in parking created at nodes where trips typically start (and are matched to previous trips ending in the neighborhood). These parking spaces can then be used for long-term parking as well in later steps of the estimation. Results from this option are presented as ``original results'' in Figs.~\ref{ratios1} and~\ref{ratios2} and have a fairly constant ratio of parking to vehicles regardless of $r_\mathrm{max}$.

	The orignal method (Algorithm~2) can be easily modified to not include the possibility for trip connections (case \#1). In this case, whenever a trip is finished, the car serving it needs to find (long-term) parking, while new trip can only be served by cars that are idle (parked). This removes the requirement that parking is available at the trip start nodes. At the same time, this can result in unrealistic dispatching, especially if cars travel to far away parking lots instead of serving passengers closer by. Also, this case has the potential of underestimating parking, since all short-term parking requirements are neglected. This case is presented in Figs.~\ref{ratios1} and~\ref{ratios2} as the solution ``without trip connections''. In this case, the ratio of parking and cars starts out as constant as well and then starts to decline above $r_\mathrm{max} = 2\,\mathrm{km}$, reaching almost one parking per car already at $r_\mathrm{max} = 10\,\mathrm{km}$. This comes at the expense of further increase in VKT (between 10\%-15\% more than in the previous case). The tradeoff between parking and VKT however stays on the same trend as in the previous case (see Fig.~\ref{ratios2}).

	An alternate way to manage parking demand is to explicitely separate parking into short-term and long-term categories as we did in the analysis considering extra wait times at the start of trips ($T_W > 0$ cases). Here, we set $T_W = 0$ (i.e.~cars normally do not have to wait), but require that parking that happens when connecting trips is accounted as short-term parking and is completely separted from long-term parking (i.e.~cannot be used for long-term parking). This is presented as the ``$T_W = 0$'' case in Figs.~\ref{ratios1} and~\ref{ratios2}, with separate plots showing only the demand for long-term parking and the combined parking demand. If we only consider long-term parking, the ratio of parking to cars again approaches 1 (but only in the $r_\mathrm{max} = $ unlimited case). If we consider combined parking demand, the ratio stays above 2 up to $r_\mathrm{max} = 5\,\mathrm{km}$ and starts to decrease after. Even for $r_\mathrm{max} = 10\,\mathrm{km}$, it is still just around $1.7$, indicating that decreasing the ratio of parking to cars below $2$ while realistically accounting for parking can only happen with high $r_\mathrm{max}$ search radius, i.e.~at the expense of large increases in VKT.
	
	We emphasize that in all cases, despite the further decrease in parking demand and increase in VKT, the relationship between these two quantities is still well explained by the exponential form found in the main text of the paper (shown as a black line in Fig.~\ref{ratios2}).

\section{Estimating current parking supply}
		
	We base the estimation of the current parking supply on minimum parking requirements and the need to find parking for each trip in the trip dataset we use. In more details, we arrived at this estimate of current parking supply with the following procedure: for housing constructed by the Housing Development Board of Singapore (HDB), we distribute the total number of parking reported officially evenly among all such apartments in Singapore. The aggregate number available from the government source~\cite{hdbparking} is 640,188, while the SimMobility database includes 1,188,649 HDB housing units in total, giving on average $0.5386$ parking spaces per unit. We note that parking managed by the HDB is often used for multiple purposes, especially for retail as some of these are strategically positioned in local town centers and offer access to the general public. For other types of development, we use the minimum parking requirements published by the Land Transport Authority (LTA) that was valid for the period before February 2019~\cite{ltaparking}. We note that LTA imposed significant changes from February 2019, lowering minimum requirements and establishing limits on maximum allowed parking for new developments~\cite{ltaparking2}. Since the buildings considered in this work were completed before this date, we used the older standard. Based on this, for private housing (i.e.~housing not constructed by the HDB), we assume one parking space per unit, giving a total of 321,974. Beside housing, we have four use categories for buildings in the SimMobility database: office, retail, factory and other with aggregate floor areas of $7.2$, $5.2$, $34.9$ and~$23.6$ million square meters respectively. For these, we use minimum parking requirements that are most appropriate, resulting in a total of 22,458 parking spaces related to office use, 24,737 parking spaces related to retail, 77,577 parking spaces related to factory use and 78,829 parking spaces for other uses. Furthermore, we have 25,740 parking spaces managed by the Urban Redevelopment Authority (URA), mostly on-street parking and some parking lots~\cite{uraparking}. Adding these up, we have a total of 1,191,503 parking spaces, giving a ratio of $1.762$ parking spaces per person for our simulated population. We display a summary of different types of parking Table~\ref{tab_parking}.
		
	The analysis above provides an approximate lower bound on parking provision, since the number of spaces for many uses is estimated based on \emph{minimum} parking requirements; limits on maximum parking in Singapore were established since February 2019, and thus do not apply to the buildings considered in our study yet~\cite{ltaparking2}. In accordance with the goverment's aim to be ``car-lite'', minimum parking requirements in Singapore are quite low, up to 5~to~10 times lower than those in use in the USA~\cite{Chester2015} for certain commercial uses. However, some developers may have voluntarily provided parking in excess of these minimums. Therefore, we adjust the distribution of parking spaces upwards based on trip data from SimMobility. We do this by performing a simplified estimate for the current trips made in private cars, where we require everyone to be able to park at the start and end location of their trips exactly (similarly to Algorithm~1 in our previous work~\cite{paper1}, but with zero search radius). We note the trip start and end locations were aggregated to a set of 4,529 discrete locations, which can be interpreted as potential locations for car parks. To combine these results with the previous estimation, we map the building locations to all possible trip start locations in a $250\,\mathrm{m}$ radius around them, and distribute the estimated parking in a way that minimizes the discrepancy between the two results. After this, for each location, we take the larger value from the two estimates as our final estimate on parking there. The main idea behind this is to adjust estimates in locations where basing them only on minimum parking requirements gives unrealistically low values; this is the case especially for office buildings in the CBD area, which is the target of a significantly high number of commuting trips. After these adjustments, we arrive at a number of 1,369,576 parking places, or $2.02$ per person in our simulation. We note that this is still a relatively low number, thus we can assume that this is a lower bound on the real number of parking in Singapore. We display the spatial distribution of parking spaces obtained with this method in Fig.~\ref{spdist}.
	
	\begin{table*}
		\centering
		\begin{tabular}{r|r|r|r}
			use type & total demand & parking requirement & total parking \\ \hline \hline
			public housing (HDB) & 1,188,649 units & N/A & 640,188 \\ \hline
			private housing & 321,974 units & 1 / unit & 321,974 \\ \hline \hline
			office (zone 1) &  4,362,915$\,\mathrm{m}^{2}$ & 1 / 450$\,\mathrm{m}^{2}$ & 9,695 \\ \hline 
			office (zone 2) &  1,590,091$\,\mathrm{m}^{2}$ & 1 / 250$\,\mathrm{m}^{2}$ & 6,360 \\ \hline 
			office (zone 3) &  1,280,454$\,\mathrm{m}^{2}$ & 1 / 200$\,\mathrm{m}^{2}$ & 6,402 \\ \hline \hline 
			retail (zone 1) &  1,657,742$\,\mathrm{m}^{2}$ & 1 / 400$\,\mathrm{m}^{2}$ & 4,144 \\ \hline
			retail (zone 2) &  1,812,934$\,\mathrm{m}^{2}$ & 1 / 200$\,\mathrm{m}^{2}$ & 9,065 \\ \hline
			retail (zone 3) &  1,729,218$\,\mathrm{m}^{2}$ & 1 / 150$\,\mathrm{m}^{2}$ & 11,528 \\ \hline \hline 
			factory & 34,909,726$\,\mathrm{m}^{2}$ & 1 / 450$\,\mathrm{m}^{2}$ & 77,577 \\ \hline
			other   & 23,648,697$\,\mathrm{m}^{2}$ & 1 / 300$\,\mathrm{m}^{2}$ & 78,829 \\ \hline
			URA parking lots & N/A & N/A & 25,740 \\ \hline \hline 
			\multicolumn{3}{l|}{Total estimated} & 1,191,502 \\ \hline
			\multicolumn{3}{l|}{Total used in trips} & 1,064,952 \\ \hline \hline
			\multicolumn{3}{l|}{Combined estimate} & 1,369,576
		\end{tabular}
		\caption{Summary of estimates of different types of parking in Singapore. Parking is given as an aggregate number for housing managed by the HDB, while it is calculated based on minimum requirements for all other types. Office and retail uses have different parking requirements in three city zones. Zone 1 is defined as the central business district and its vicinity, zone 2 is defined as the area within $400\,\mathrm{m}$ distance from any rapid transit station, while zone 3 is the rest of the city. We used the ``retail'' category for any commercial establishment, while the minimum parking requirement for the ``factory'' and ``other'' categories were determined as the average of multiple relevant categories.}
		\label{tab_parking}
	\end{table*}

	\section{Spatial distribution of parking demand}

	\begin{figure*}
		\centering
		\includegraphics[width=3.5in]{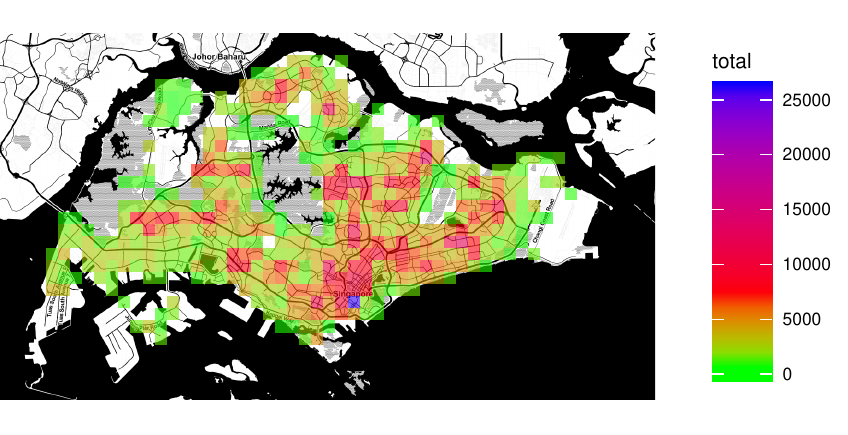}
		\caption{Spatial distribution of parking spaces as estimated from the SimMobility building and trip database. The total number of parking spaces is 1.37~million. Resolution of the grid displayed is $1\,\mathrm{km} \times 1\,\mathrm{km}$ similarly to Fig.~\ref{spdist1}, i.e.~the numbers given are parking per square kilometer. Note that the color scale is different from Fig.~\ref{spdist1}.}
		\label{spdist}
	\end{figure*}
	
	\begin{figure*}
		\centering
		\vspace{3ex}
		\begin{overpic}{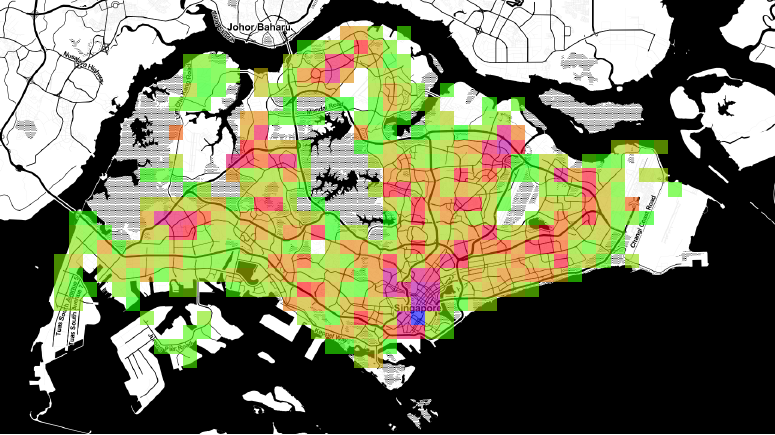}  
			\put(50,60){\makebox(0,0){$r_\mathrm{max} = 500\,\mathrm{m}$}}
		\end{overpic}
		\quad
		\begin{overpic}{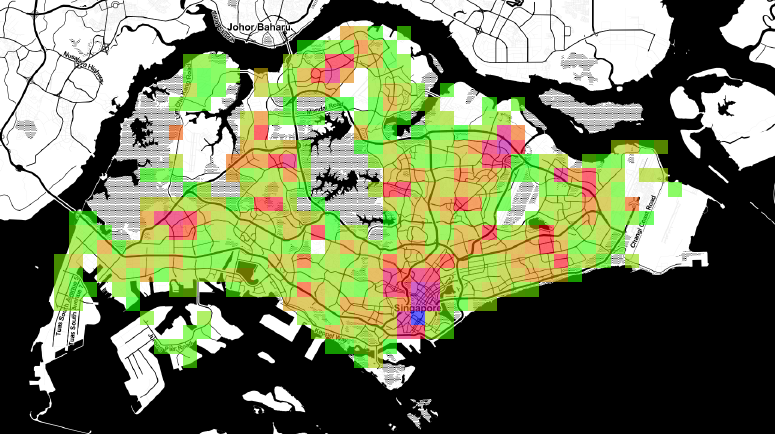} 
			\put(100,0){\includegraphics[width=1.8cm]{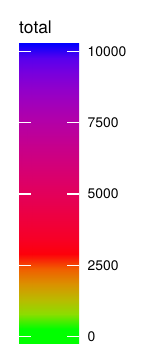}}
			\put(50,60){\makebox(0,0){$r_\mathrm{max} = 1\,\mathrm{km}$}}
		\end{overpic} \\[5ex]
		\begin{overpic}{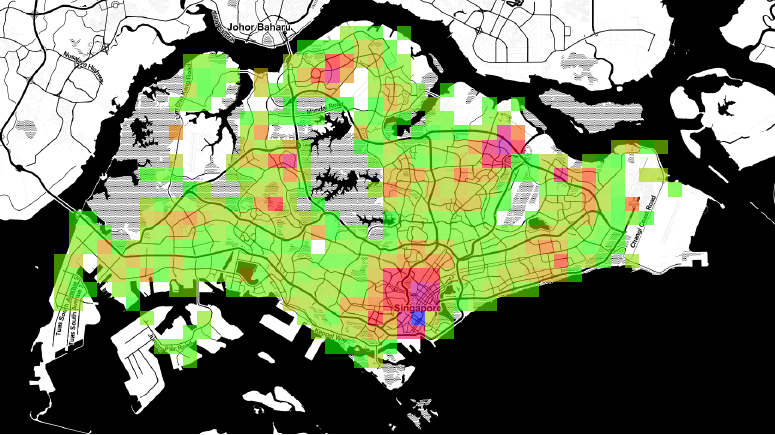}
			\put(50,60){\makebox(0,0){$r_\mathrm{max} = 2\,\mathrm{km}$}}
		\end{overpic} \quad
		\begin{overpic}{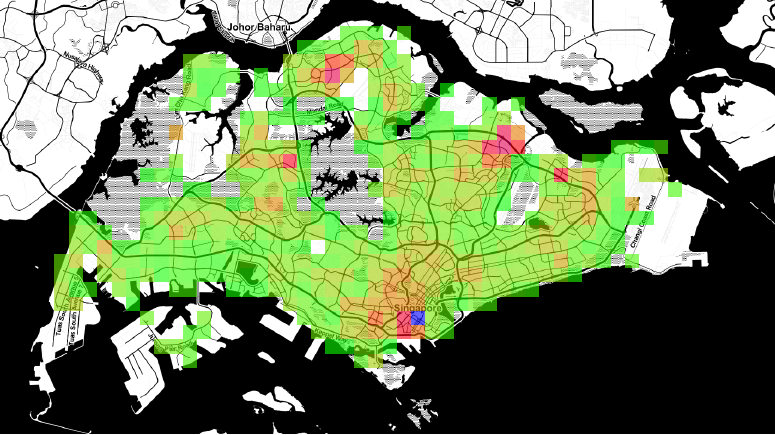}
			\put(50,60){\makebox(0,0){$r_\mathrm{max} = 5\,\mathrm{km}$}}
		\end{overpic} \\[5ex]
		\begin{overpic}{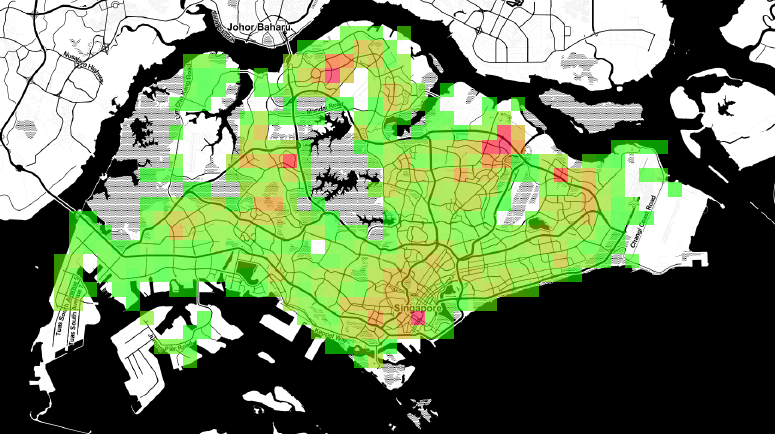}
			\put(50,60){\makebox(0,0){$r_\mathrm{max} = 10\,\mathrm{km}$}}
		\end{overpic} \quad
		\begin{overpic}{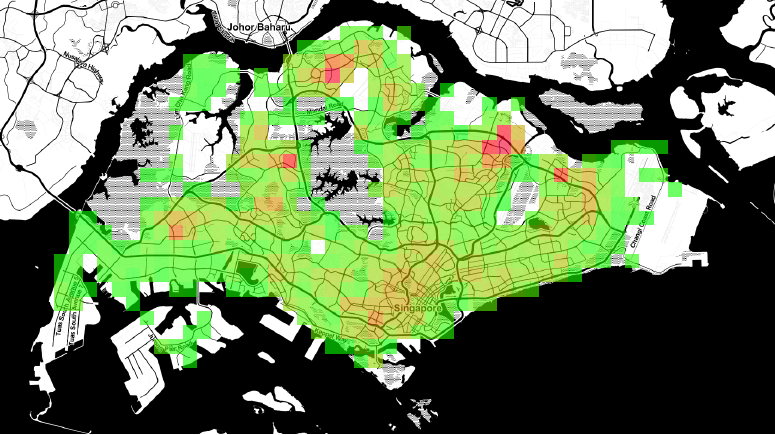}
			\put(50,60){\makebox(0,0){$r_\mathrm{max} = $ unlimited}}
		\end{overpic}
		\caption{Spatial distribution of parking demand found as a result of the batched bipartite matching estimation, i.e.~Algorithm~\ref{algsteps}. Resolution of the grid displayed is $1\,\mathrm{km} \times 1\,\mathrm{km}$, i.e.~the number of parking spaces are given as per square kilometer. Results are displayed for $r_\mathrm{max}$ values between $500\,\mathrm{m}$ and unlimited. Note that the color scale on all panels is the same, ranging from 0 to 10,000 parking spaces per square kilometer. For $r_\mathrm{max} \leq 5\,\mathrm{km}$, the grid cell with the most parking demand in the central business district is an outlier, having 24,723, 24,313, 21,326 and 11,577 parking spaces for $r_\mathrm{max} = 500\,\mathrm{m}$, $1\,\mathrm{km}$, $2\,\mathrm{km}$ and $5\,\mathrm{km}$ respectively (the color scale is saturated at 10,000 parking spaces to faciliate the comparison of results in each case).}
		\label{spdist1}
	\end{figure*}
	
\begin{figure*}
		\centering
		\vspace{3ex}
		\begin{overpic}{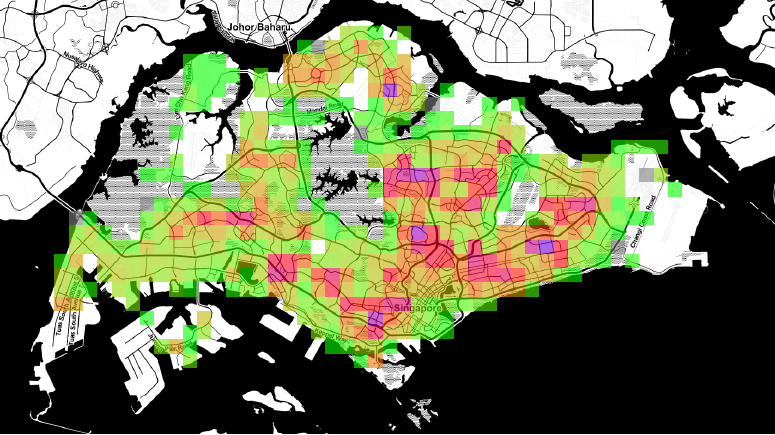}  
			\put(50,60){\makebox(0,0){$r_\mathrm{max} = 500\,\mathrm{m}$}}
		\end{overpic}
		\quad
		\begin{overpic}{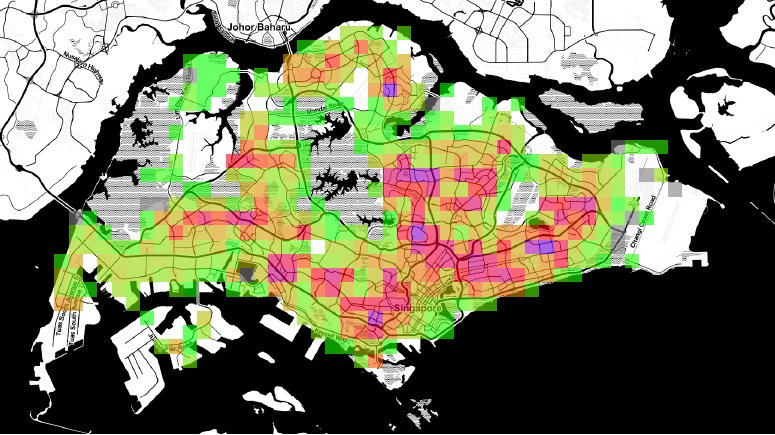} 
			\put(100,0){\includegraphics[width=1.8cm]{parking_dist_new_legend}}
			\put(50,60){\makebox(0,0){$r_\mathrm{max} = 1\,\mathrm{km}$}}
		\end{overpic} \\[5ex]
		\begin{overpic}{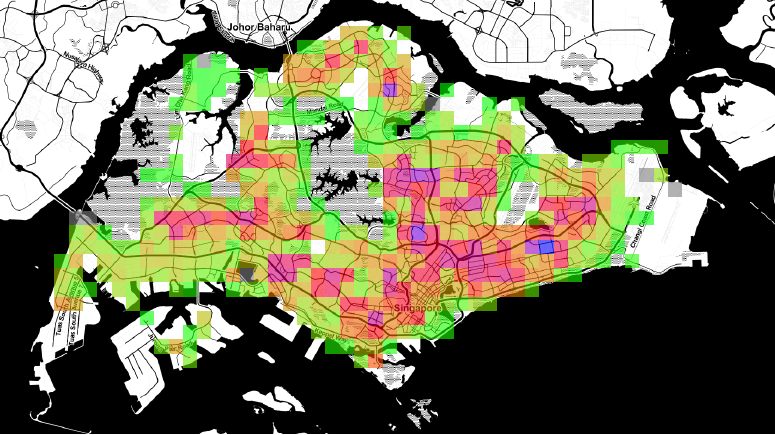}
			\put(50,60){\makebox(0,0){$r_\mathrm{max} = 2\,\mathrm{km}$}}
		\end{overpic} \quad
		\begin{overpic}{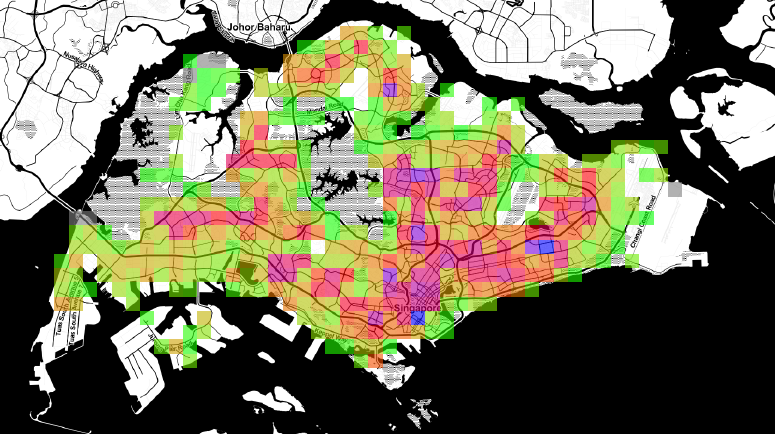}
			\put(50,60){\makebox(0,0){$r_\mathrm{max} = 5\,\mathrm{km}$}}
		\end{overpic} \\[5ex]
		\begin{overpic}{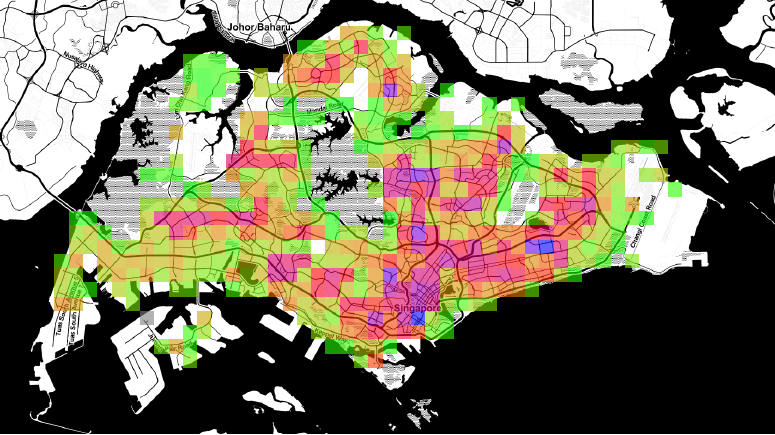}
			\put(50,60){\makebox(0,0){$r_\mathrm{max} = 10\,\mathrm{km}$}}
		\end{overpic} \quad
		\begin{overpic}{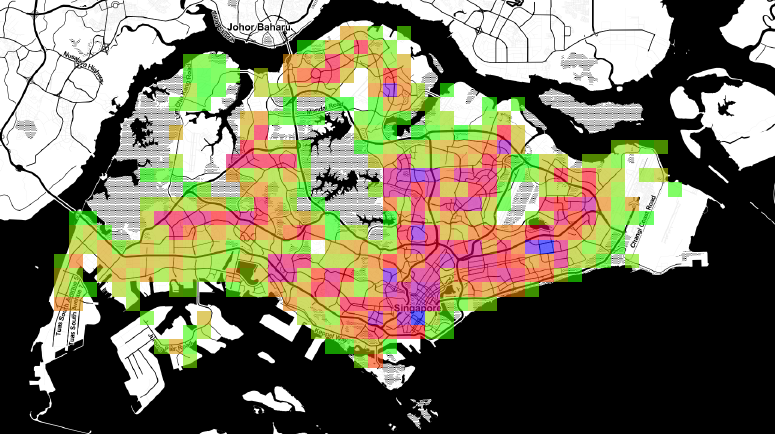}
			\put(50,60){\makebox(0,0){$r_\mathrm{max} = $ unlimited}}
		\end{overpic}
		\caption{Spatial distribution of savings in parking demand, i.e.~the difference between the spatial distributions displayed in Figs.~\ref{spdist1} and~\ref{spdist}.}
		\label{spdiff}
	\end{figure*}

	Our analysis allows us to investigate not only fleet size and aggregate parking needs, but also the spatial distribution of current parking supply and future demand if OVs are adopted. We display these distributions in Fig.~\ref{spdist} (estimation of current supply) and~\ref{spdist1} (estimation parking dmenad of OVs). Furthermore, we display the difference between the two (i.e.~the possible savings due to adoption of OVs) in Fig.~\ref{spdiff}. The main difference is that while the current distribution of parking spaces is quite spread out over the dense areas of Singapore, with significant parking provisions for residential areas, the parking demand of the OV fleet is highly concentrated in the CBD, especially for low values of $r_\mathrm{max}$ with a few smaller peaks in residential centers in Punggol, Woodlands, and Pasir Ris.
	
	A main contributing factor to this spatial asymmetry is the high concentration of jobs in the CBD, thus the imbalance of commuting flows going there. Increasing $r_\mathrm{max}$ results in a significant decrease in the parking demand in the city center in our estimation; this is understandable, since a higher $r_\mathrm{max}$ allows OVs bringing workers to the centre to choose a parking location during the day in a much larger area.
	On the other hand, in residential areas, there is a potential oversupply of parking even in the current situation, with a total of $962$~thousand parking spaces currently supplied for primarily residential use, or $1.42$ residential parking spaces per private car user. We note that many of these parking spaces are actually usable by the general public, i.e.~parking managed by the Housing Development Board (HDB) of Singapore.
	
	Looking at the differences in Fig.~\ref{spdiff}, we see that most areas can be expected to experience a significant decrease of parking demand; this is most prominent in dense residential areas such as Bedok, Toa Payoh, Ang Mo Kio or Clementi, where the decrease in demand is already significant for low $r_\mathrm{max}$ values. A significant amount of parking in these settings is provided by the HDB as surface parking lots and separate parking garages, thus repurposing the land area taken up by these could be more easily achieved than underground parking facilities found with many commercial and private residential buildings.
	
\end{document}